\documentclass[10pt,journal,compsoc]{IEEEtran}
\usepackage{subfigure}
\usepackage{amsmath,amsfonts}
\usepackage{algorithmic}
\usepackage{algorithm}
\usepackage{array}
\usepackage[caption=false,font=normalsize,labelfont=sf,textfont=sf]{subfig}
\usepackage{textcomp}
\usepackage{stfloats}
\usepackage{url}
\usepackage{verbatim}
\usepackage{graphicx}
\usepackage{cite}
\usepackage{color}
\usepackage{booktabs}
\usepackage{threeparttable}
\usepackage{bbding}
\usepackage{pifont}
\usepackage{multicol}

\ifodd 1
\newcommand{\com}[1]{\textbf{\color{red}(COMMENT: #1)}} 
\else

\newcommand{\com}[1]{}
\fi

\newcommand{\revise}[1]{{\color{black}{#1}}}
\newcommand{\majorrevise}[1]{{\color{black}{#1}}}

\def\fig{Fig.}

\def\eg{e.g.}

\ifCLASSINFOpdf
\else
\fi
\usepackage{balance}

\begin{document}

%
\title{\revise{Aerial Shepherds: Enabling Hierarchical Localization in Heterogeneous MAV Swarms}}





%
%
%

\author{
Haoyang Wang, 
Jingao Xu,
Chenyu Zhao,
Yuhan Cheng,
Xuecheng Chen,
Chaopeng Hong,\\
Xiao-Ping Zhang, ~\IEEEmembership{Fellow, ~IEEE},
Yunhao Liu, ~\IEEEmembership{Fellow, ~IEEE},
Xinlei Chen, ~\IEEEmembership{Member, ~IEEE}

\IEEEcompsocitemizethanks{
\IEEEcompsocthanksitem A preliminary version of this article appeared in IEEE International Conference on Computer Communications (IEEE INFOCOM 2024) \cite{wang2024transformloc}.

\IEEEcompsocthanksitem Haoyang Wang, Chenyu Zhao, Yuhan Cheng and Xuecheng Chen are with Tsinghua-Berkeley Shenzhen Institute, Shenzhen International Graduate School, Tsinghua University, China. \\
E-mail: \{wanghaoyang0428, zhaocyhi, yhhncc, 1243566go\}@gmail.com

\IEEEcompsocthanksitem Jingao Xu and Yunhao Liu are with the School of Software and BNRist, Tsinghua University, Beijing 100084, China. \\
Email: \{xujingao13, yunhaoliu\}@gmail.com



\IEEEcompsocthanksitem Chaopeng Hong and Xiao-Ping Zhang are with Shenzhen International Graduate School, Tsinghua University, China.\\
E-mail: \{hongcp, xiaoping.zhang\}@sz.tsinghua.edu.cn

\IEEEcompsocthanksitem Xinlei Chen is with Shenzhen International Graduate School, Tsinghua University, China, Pengcheng Laboratory, Shenzhen, China, and RISC-V International Open Source Laboratory, Shenzhen, China. \\
E-mail: chen.xinlei@sz.tsinghua.edu.cn

\IEEEcompsocthanksitem Corresponding authors: Xinlei Chen.
\IEEEcompsocthanksitem Manuscript submitted March 2024.

}
}

%
%

\markboth{IEEE TRANSACTIONS ON MOBILE COMPUTING}%
{Wang. H \MakeLowercase{\textit{et al.}}: TransformLoc}
%



\IEEEtitleabstractindextext{%
\begin{abstract}

\revise{A heterogeneous micro aerial vehicles (MAV) swarm consists of resource-intensive but expensive advanced MAVs (AMAVs) and resource-limited but cost-effective basic MAVs (BMAVs), offering opportunities in diverse fields. }
Accurate and real-time localization is crucial for MAV swarms, but current practices lack a low-cost, high-precision, and real-time solution, especially for lightweight BMAVs. 
We find an opportunity to accomplish the task by transforming AMAVs into mobile localization infrastructures for BMAVs. 
\revise{However, translating this insight into a practical system is challenging due to issues in estimating locations with diverse and unknown localization errors of BMAVs, and allocating resources of AMAVs considering interconnected influential factors.}
This work introduces TransformLoc, a new framework that transforms AMAVs into mobile localization infrastructures, specifically designed for low-cost and resource-constrained BMAVs. 
\revise{We design an error-aware joint location estimation model to perform intermittent joint estimation for BMAVs and introduce a similarity-instructed adaptive grouping-scheduling strategy to allocate resources of AMAVs dynamically.}
TransformLoc achieves a collaborative, adaptive, and cost-effective localization system suitable for large-scale heterogeneous MAV swarms. 
\revise{We implement and validate TransformLoc on industrial drones. Results show it outperforms all baselines by up to 68\% in localization performance, improving navigation success rates by 60\%. Extensive robustness and ablation experiments further highlight the superiority of its design.}
\end{abstract}

}

\maketitle

\IEEEdisplaynontitleabstractindextext

%
\IEEEpeerreviewmaketitle

\vspace{-2cm}
\IEEEraisesectionheading{\section{Introduction}}


\IEEEPARstart{H}{eterogeneous} micro aerial vehicle (MAV) swarms demonstrate transformative potential in addressing 4D (deep, dull, dangerous, dirty) tasks, particularly in scenarios requiring prompt responses, such as search and rescue during diverse disasters \cite{rashid2020socialdrone}, \majorrevise{gas leak source detection \cite{duisterhof2021sniffy, francis2022gas, chen2019odor}}, and wildfire suppression \cite{khochare2021heuristic}. 
\majorrevise{These swarms capitalize on inherent advantages, including working scalability \cite{bertizzolo2020swarmcontrol}, flexibility \cite{wang2021lifesaving}, and adaptability \cite{li2021physical}.
Forecasts project that the market size for MAV swarm-supported applications will reach \$279 billion by 2032 \cite{Global-Drones}.}

\begin{figure}[t]
    \centering
        \includegraphics[width=0.95\columnwidth]{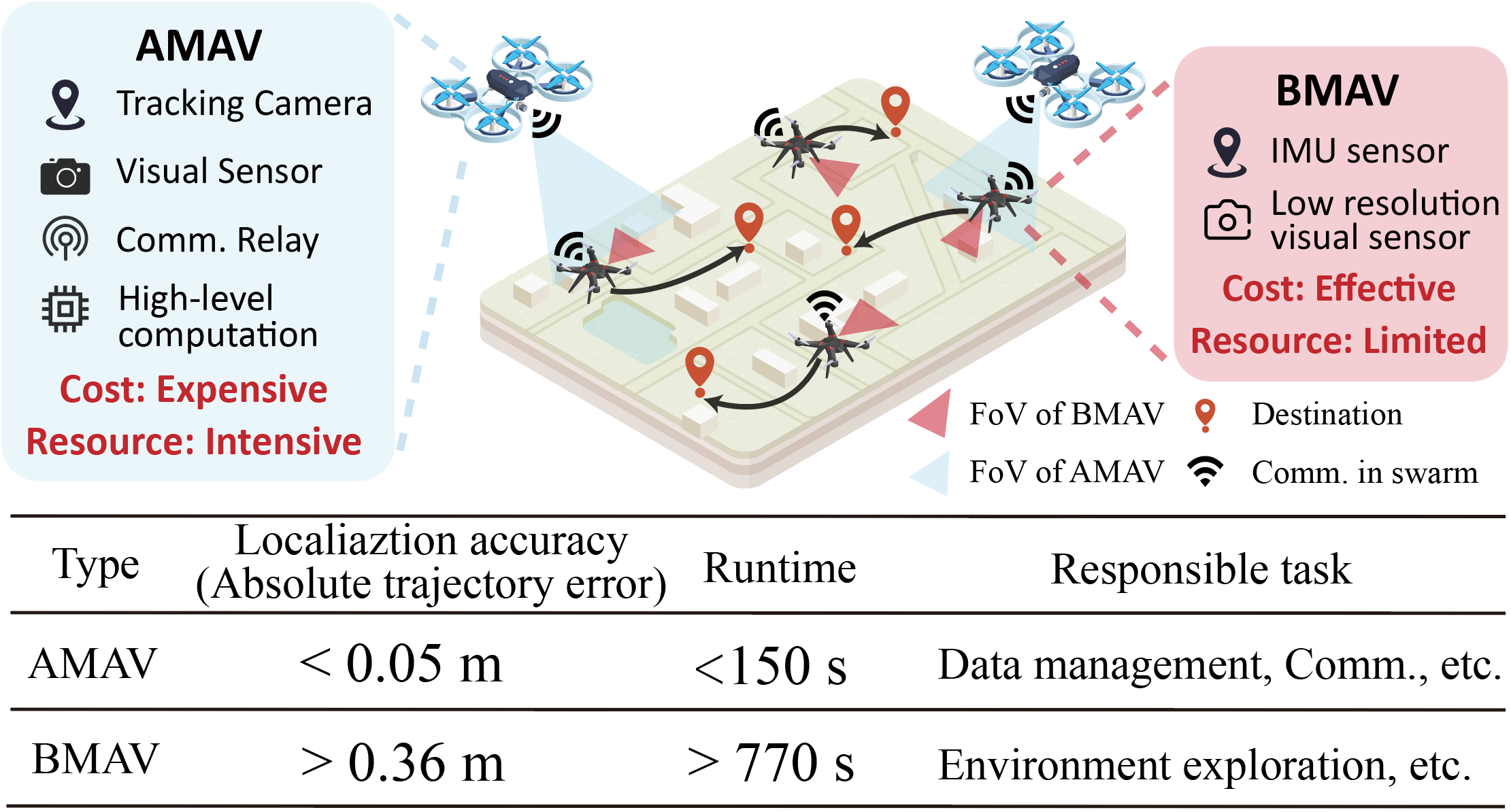}
    \vspace{-0.2cm}
     \caption{Introduction of a heterogeneous MAV swarm composed of two types: AMAVs and BMAVs. \revise{AMAVs are resource-intensive, offering lower localization error and reduced latency, whereas BMAVs, with their resource limitations, experience higher error and latency \protect\footnotemark[1].}}
    \label{intro}
    \vspace{-0.5cm}
\end{figure}

A heterogeneous MAV swarm typically comprises: 
$(i)$ a limited number of advanced MAVs (AMAVs) endowed with intensive capabilities (\eg, sensing, computing, etc.) albeit at a high cost \cite{shen2022joint};
and $(ii)$ a larger cohort of resource-limited yet cost-effective basic MAVs (BMAVs) \cite{wang2022h}, as illustrated in \fig.~\ref{intro}.
In a heterogeneous MAV swarm, the AMAV primarily manages data, communication, and intricate computational tasks, while the latter is dispersed into (hazardous) target areas for data collection and environmental exploration \cite{li2022intelligent}. 
\revise{The collaboration paradigm between AMAVs and BMAVs achieves an optimal balance between overall capabilities and cost, facilitating its widespread adoption in extensive applications, including Amazon disaster relief \cite{rashid2020computational}, DJI industrial inspection \cite{yang2019intelli}, aerial imaging \cite{trotta2018uavs}.}

\footnotetext[1]{Fig. 1 and Fig.2a present results obtained using MH\_02 from the EuRoC dataset \cite{burri2016euroc} and CCM-SLAM \cite{schmuck2019ccm}. The AMAV is equipped with an Intel(R) i7-8750H processor and 32GB RAM, while the BMAV is equipped with a Cortex-A53 processor and 1GB RAM.}

Nevertheless, the disparate onboard resources result in imbalanced localization performance between AMAVs and BMAVs, a critical capability for flight control \cite{zhao2023smoothlander}, obstacle avoidance \cite{xu2020edge}, and more. 
Specifically, the BMAV experiences rapid accumulation of localization errors due to its noisy sensing data and limited computational capability, impeding its ability to achieve precise and real-time localization comparable to the AMAV.
\revise{As illustrated in \fig~\ref{intro}, for a same localization and navigation task, the AMAV completes it in \textless150 seconds with a localization error of \textless0.05 meters. 
In contrast, the BMAV takes \textgreater770 seconds and has a localization error of \textgreater0.36 meters.
This stark difference in sensing and computation capabilities results in significantly lower localization efficiency for the BMAV compared to the AMAV.}

Regrettably, existing approaches fail to provide viable solutions for BMAV localization in quick response scenarios, which can be divided into two distinct types:\\
\noindent $\bullet$ \textbf{Extra infrastructure-based solutions}. 
These approaches leverage external devices to furnish reference signals, such as GPS \cite{sharp2022authentication, gowda2016tracking}, RTK \cite{wang2022micnest}, radio \cite{li2021infocom_itoloc, chi2022wi}, or acoustic \cite{wang2022micnest}, for MAV localization. They necessitate: $(i)$ deploying expensive localization infrastructure in the operation site, often in a dense configuration, and $(ii)$ maintaining a line-of-sight connection between the localization infrastructure and MAVs. 
However, in quick response scenarios like urban disaster relief, fulfilling either requirement becomes challenging, resulting in localization failures \cite{chen2015drunkwalk}.

\begin{figure}[t]
\setlength{\abovecaptionskip}{0.cm} 
\setlength{\belowcaptionskip}{-0.4cm} 
\setlength{\subfigcapskip}{-0.1cm}  
\centering
    \subfigure[Accuracy \& latency on BMAV]{
        \centering
            \includegraphics[width=0.45\columnwidth]{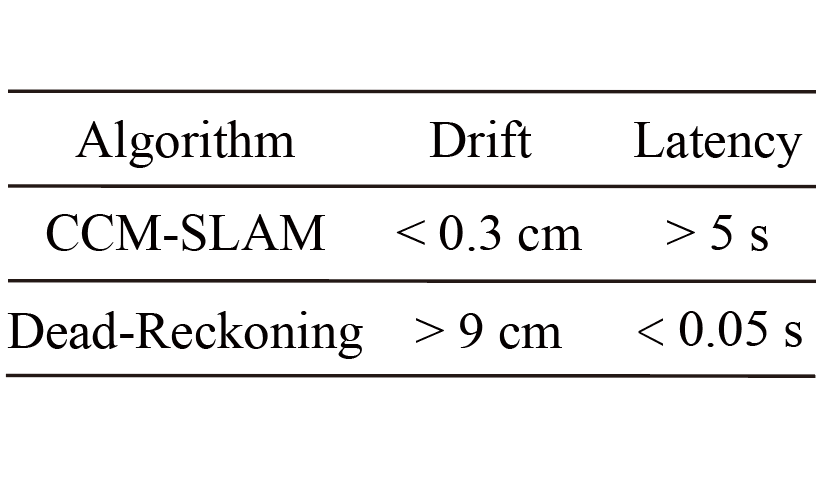}
    }
    \subfigure[Accuracy of AMAV observation]{
        \centering
            \includegraphics[width=0.48\columnwidth]{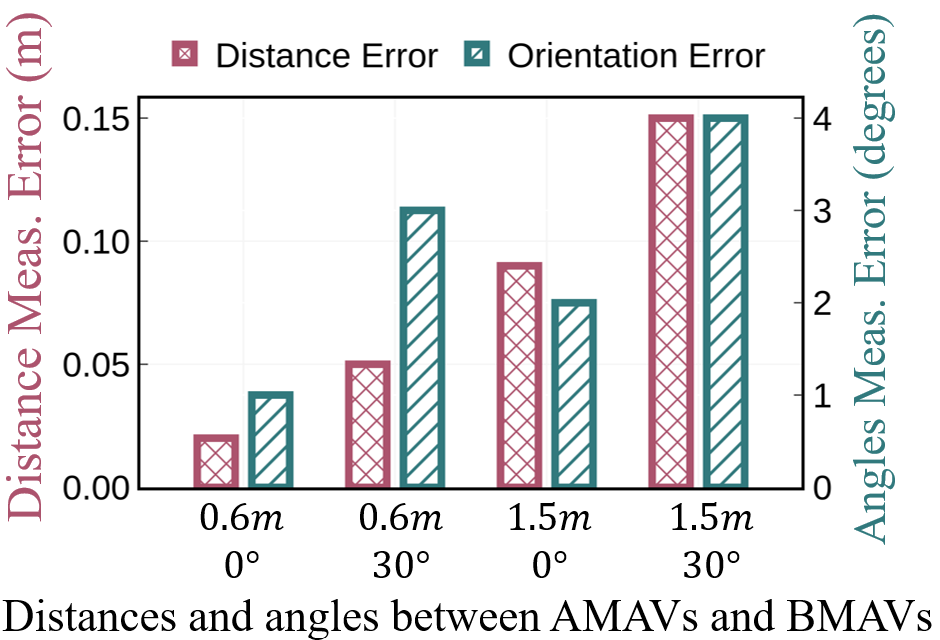}
    }
\caption{Motivating example. (a) BMAVs contend with constrained onboard resources, necessitating a balance between accuracy and latency. (b) The precision of observations is influenced by the distances and angles between MAVs.}
\label{motivation}
\vspace{-0.4cm}
\end{figure}

\noindent $\bullet$ \textbf{Intra on-board sensor-based solutions.}
These approaches utilize onboard sensors, including cameras \cite{xu2022swarmmap},  spectral camera\cite{fu2021coded, li2024latent}, IMU \cite{hu2023combining}, LiDAR \cite{li2023leovr}, and Radar \cite{lu2020see}, in conjunction with simultaneous localization and mapping (SLAM) techniques, to achieve precise location estimation \cite{lu2020milliego, dong2019infocom_pairnavi}. 
Moreover, leveraging on image enhancement, these methods can adapt to complex environments \cite{fu2022low, zhang2024deep, wei2021physics}.
These methods demand substantial on-board sensing and computing capabilities, rendering them suitable only for AMAVs.
Conversely, the limited on-board resources of BMAVs result in computation delays or considerable error accumulation \cite{cao2022edge}.
\revise{As depicted in \fig~\ref{motivation}a, we tested different localization algorithms on BMAVs. 
When employing CCM-SLAM, the BMAV achieved a low localization error (\textgreater0.3 $cm$) but required a long runtime (\textgreater5 s). 
In contrast, when using Dead-Reckoning, the error increased (\textgreater9 $cm$), but the latency is significantly reduced (\textless0.05 s).
Therefore, BMAVs face limited onboard resources, necessitating a balance between latency and accuracy.}


\revise{Therefore, this paper aims to enhance the localization accuracy and efficiency of BMAVs, leveraging limited onboard sensing and computing capabilities without dependence on pre-deployed localization infrastructures. 
The \textbf{key insight} involves repurposing a select number of AMAVs as aerial shepherds, functioning as \textit{mobile localization infrastructures}, which redistribute their sensing and computing capabilities to enhance the location estimation of a larger fleet of BMAVs.
Specifically, AMAVs contribute external observations (via visual sensors) to rectify cumulative location estimation errors \cite{wang2016apriltag,xu2021mobisys_followupar}. 
In addition, to optimize real-time performance, most intricate computational tasks should be primarily executed within the AMAVs.}

\majorrevise{However, converting this insight into a functional system presents significant complexity, as it necessitates the resolution of two pivotal technical challenges:}

\noindent \textbf{$\bullet$(C1) Unknown \& Diverse localization errors of BMAVs.} 
On one hand, diverse factors such as sensing noises and varying operational conditions contribute to the accumulation of location estimation errors among BMAVs.
On the other hand, the limited field of view (FoV) of each AMAV restricts its ability to serve as localization infrastructure for only a subset of BMAVs. 
Therefore, AMAVs ought to furnish external observations to aid in correcting estimation errors of BMAVs exhibiting significant deviations.
However, the real-time derivation of BMAVs' localization errors presents challenges due to the absence of static location references.
Consequently, without timely external observations from AMAVs, localization errors of BMAVs may persistently accumulate. 
Therefore, it is crucial to address how to correct BMAVs' locations with AMAVs' observations.

\noindent \textbf{$\bullet$(C2) Resource allocation given coupled influential factors.}
Optimizing the allocation of sensing and computing resources from a limited number of AMAVs to a larger fleet of BMAVs as localization infrastructure entails optimization in a high-dimensional decision space.
Firstly, even if the initial constraint (\textbf{C1}) is resolved, the involvement of multiple AMAVs leads to an exponential growth in search space. 
Secondly, given the perpetual motion of both AMAVs and BMAVs, the set of observable BMAVs for each AMAV constantly shifts, demanding dynamic adjustments in resource allocation strategies. 
Thirdly, the accuracy of AMAVs' observations, influenced by varying distances and bearing angles between AMAVs and BMAVs, further complicates the search space for resource allocation, directly impacting localization effectiveness. 
\majorrevise{As illustrated in \fig~\ref{motivation}b, we conduct measurement experiments with a camera and AprilTag.} 
At a short distance and small angle (0.6 meters and 0 degrees), the observed distance and angle errors are minimal (0.02 meters and 1 degree). However, as the distance or angle increases, the errors grow significantly (at 1.5 meters and 30 degrees, the errors reach 0.15 meters and 4 degrees).
These intertwined factors render resource allocation exceedingly intricate for implementation on MAVs.

\textbf{Our work.} To address these challenges, this paper introduces \textit{TransformLoc}, an innovative collaborative and adaptive location estimation framework designed for heterogeneous MAV swarms.
TransformLoc dynamically converts AMAVs, equipped with intensive onboard capabilities, into aerial shepherds and mobile localization infrastructures to enhance location estimation for resource-constrained BMAVs.
By concentrating intensive sensing and computing resources on a select few AMAVs, TransformLoc ensures accurate and real-time localization for the entire swarm, eliminating the need for external localization infrastructures. 
As a result, this framework promotes the extensive adoption of swarm technology.

To tackle \textbf{C1}, we devise an \textit{error-aware joint location estimation} model. This model incorporates an uncertainty-aided inference method, allowing AMAVs to prioritize providing observations to BMAVs with greater errors. Subsequently, it integrates inaccurate estimations from BMAVs with intermittent external observations from AMAVs to facilitate periodic joint location estimation for BMAVs.

\revise{To address \textbf{C2}, we introduce a \textit{similarity-instructed adaptive grouping-scheduling} strategy. 
Initially, MAVs are dynamically grouped in accordance with the principle of proximity, transforming the many-to-many resource allocation challenge into multiple one-to-many resource allocation tasks. 
Then, employing a several-step lookahead approach for BMAVs, each AMAV is scheduled in a non-myopic manner, optimizing both distance and angle. 
Concurrently, the strategy utilizes a similarity-instructed method to identify and eliminate redundant search space, retaining a smaller subset of decision space with significant distinctions. 
Ultimately, complexity is diminished from exponential to linear scale.}


We assess the performance of TransformLoc and benchmark it against baselines including the state-of-the-art (SOTA) method, using comprehensive experiments conducted on both a real-time testbed and physical feature-based simulations. 
Our findings reveal that TransformLoc achieves an average localization error of under $1m$ for BMAVs with limited onboard resources in real-time, outperforming baselines by up to $68\%$. This enhancement translates to a potential increase of up to $60\%$ in navigation success rates over baselines.
\revise{Additionally, to explore the boundaries of TransformLoc, we conduct extensive robustness evaluations, testing the system's resilience to varying numbers of MAVs and measurement noise. 
Furthermore, we perform ablation experiments, which underscored the superiority of TransformLoc's design through comprehensive results.}

\begin{figure*}[t]
    \centering
        \includegraphics[width=2.0\columnwidth]{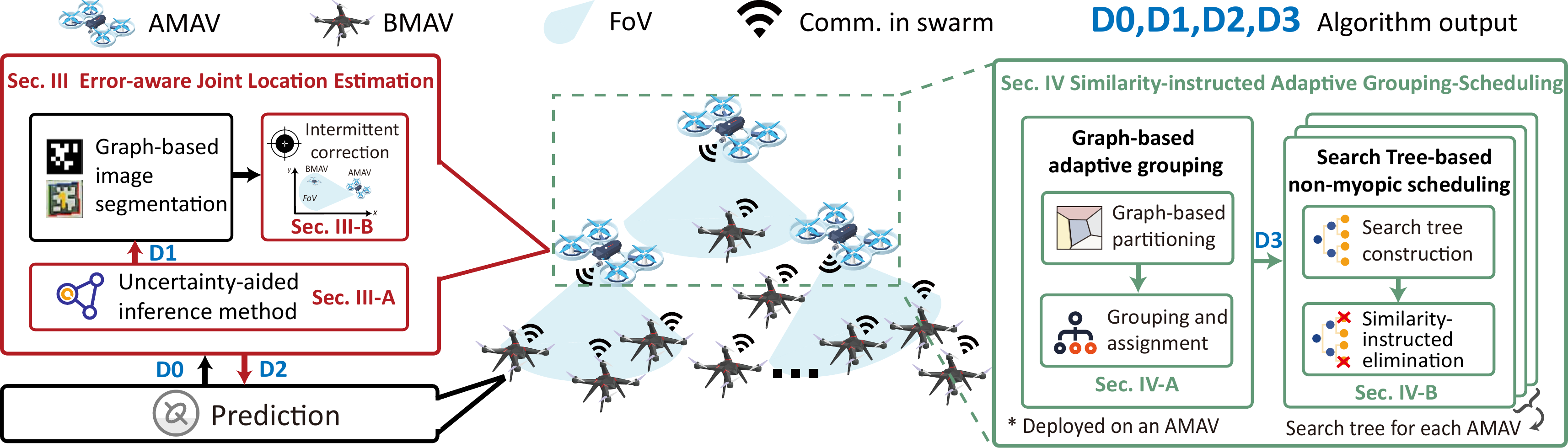}
        \vspace{-0.2cm}
    \caption{
    \revise{Demonstration of the TransformLoc framework. Initially, BMAVs estimate locations amidst noisy measurements. Leveraging an uncertainty-aided inference method, AMAV generates intermittent observations to facilitate intermittent joint location estimation for BMAVs. Following this, TransformLoc optimizes resource allocation for AMAVs through adaptive grouping and scheduling. This entails dynamically grouping MAVs and subsequently scheduling AMAVs in a non-myopic fashion, followed by a similarity-instructed elimination method to identify and eliminate redundant search space.}
    }
    \label{architecture}
    \vspace{-0.4cm}
\end{figure*} 

\noindent \textbf{Summary.} The primary contributions of this work include:

\noindent $\bullet$ We introduce TransformLoc, a novel framework that dynamically converts AMAVs into aerial shepherds and mobile localization infrastructures, improving localization accuracy and real-time performance for lightweight BMAVs.

\noindent $\bullet$ We develop an \textit{error-aware joint location estimation} model aimed at enhancing the accuracy of location estimation for BMAVs utilizing intermittent observations from AMAVs.

\noindent $\bullet$ 
\revise{We formulate a \textit{similarity-instructed adaptive grouping-scheduling} strategy to decouple the intertwined resource allocation challenge influenced by various coupled factors.}

\noindent $\bullet$ 
We verify the effectiveness of our solution through real-world experiments conducted on a real heterogeneous MAV swarm and extensive physical feature-based simulations.






The subsequent sections of the paper are structured as follows: 
Sec. \ref{3} presents an overview of TransformLoc, followed by detailed descriptions of the error-aware joint location estimation model in Sec. \ref{4} and the similarity-instructed adaptive grouping-scheduling strategy in Sec. \ref{5}. 
\majorrevise{The implementation and evaluation of TransformLoc are showcased in Sec. \ref{6}, while Sec. \ref{7}, Sec. \ref{8} and \ref{9} delve into related works, discussion, and future work of the framework, respectively, concluding the TransformLoc in Sec. \ref{10}. }

\vspace{-0.5cm}
\section{Overview}\label{3}

\subsection{Framework goal}
From a top-level perspective, TransformLoc is designed to convert AMAVs into mobile localization infrastructures tailored for BMAVs. This specialized adaptation is meticulously designed to accommodate the economic and resource limitations inherent to BMAVs, thus effectively overcoming the challenges posed by these constraints.
The framework design aims to address the following questions:

\noindent $\bullet$ \textit{How can we enhance the location estimation accuracy of BMAVs despite diverse and unknown localization errors?} TransformLoc is tasked with inferring the localization errors of BMAVs and leveraging observations provided by AMAVs to aid in localization error reduction effectively.

\noindent $\bullet$ \textit{How should we allocate sensing resources of AMAVs, considering intertwined influential factors to support BMAVs?} TransformLoc should disentangle the resource allocation challenge and guide AMAVs in a non-myopic manner, ensuring minimal overall localization errors for BMAVs.

\subsection{Framework overview}

The architecture of TransformLoc is depicted in \fig~\ref{architecture}. It comprises two primary components: the \textit{Error-aware Joint Location Estimation} model and the \textit{Similarity-instructed Adaptive Grouping-Scheduling} strategy. In the following, we will provide an overview of these two components:

\noindent $\bullet$ \textit{Error-aware Joint Location Estimation} model. 
In the initial step, BMAVs employ the state model and motion commands to generate predictions, transmitting the resulting prior distribution of location to AMAVs (\textit{D0} in \fig~\ref{architecture}).
Subsequently, the uncertainty-aided inference method (Sec. \ref{3.2}) discerns instances where BMAVs require assistance (\textit{D1} in \fig~\ref{architecture}).
Leveraging observations from the visual fiducial system, AMAVs collaboratively refine the location estimations of BMAVs requiring assistance (Sec. \ref{3.3}), transmitting the resulting posterior distribution of BMAVs' locations and motion commands back to BMAVs (\textit{D2} in \fig~ \ref{architecture}).

\revise{\noindent $\bullet$ \textit{Similarity-instructed Adaptive Grouping-Scheduling} strategy.
With a limited number of AMAVs, this strategy is tasked with allocating sensing resources from AMAVs to BMAVs. 
Initially, this strategy dynamically groups MAVs using a graph-based approach based on the spatial relationships among AMAVs (Sec. \ref4.2). 
Leveraging the grouping outcomes (\textit{D3} in \fig~ \ref{architecture}), TransformLoc then orchestrates the scheduling of each AMAV in a non-myopic manner by constructing search trees for individual AMAVs. 
To enhance efficiency of the search tree, TransformLoc further employs a similarity-instructed method to identify and eliminate redundant nodes in search trees (Sec. \ref{4.3}). 
Ultimately, TransformLoc generates motion commands for AMAVs to effectively distribute resources to BMAVs.}

\vspace{-0.2cm}
\section{Error-aware joint location estimation}\label{4}

The BMAVs present a challenge due to their diverse and unknown localization errors, complicating joint location estimation with AMAV assistance. 
In this section, we introduce an error-aware joint location estimation model rooted in the Kalman filter, focusing on localization error inference and joint location estimation for BMAVs.

\revise{Initially, we delineate our key definitions, which encompasses the state model of both BMAVs and AMAVs, as well as the observation model of AMAVs (Sec. \ref{3.1}). }
Subsequently, to facilitate AMAVs in providing observations for BMAVs with higher errors, we introduce a measure of uncertainty to reflect the accuracy of BMAVs' location estimation (Sec. \ref{3.2}).
Concurrently, we undertake two interconnected operations for BMAV location estimation (Sec. \ref{3.3}): 
(1) \textit{Prediction from motion model}, which utilizes noisy motion measurements and the state model of BMAVs to forecast location estimation.
(2) \textit{Correction from observation}, incorporating linearized observations generated by AMAVs to refine the location estimation of the BMAVs.

\revise{
\subsection{Key definitions} \label{3.1}
\subsubsection{Environmental Description} 
Consider $\Omega$ as a bounded space within $\mathbb{R}^3$ with dimensions defined by length ($L$), width ($W$), and height ($H$), where the heterogeneous MAV swarm operates. 
This swarm consists of a predetermined quantity of BMAVs and AMAVs, which operates autonomously. 
For the sake of simplicity, we assume that both types of MAVs operate at an equal altitude throughout the paper.

\subsubsection{State model of AMAV} 
The heterogeneous MAV swarm comprises $M$ AMAVs ($M > 1$). At time $t$, the state of each AMAV $A_j$ encompasses both its location and motion command. The location is represented as $\boldsymbol{x_{j,t}} = (x_{j,t}^1, x_{j,t}^2, \phi_{j,t})$, where $\phi_{j,t}$ denotes the orientation angle, and the distance between two locations is denoted as $d_{\chi}$. Regarding the motion command, $\boldsymbol{u_{j, t}} = (u_{j,t}, \omega_{j,t})$, where $u_{j,t}$ and $\omega_{j,t}$ represent translational and rotational velocities, respectively.
The AMAVs achieve accurate location estimation through advanced sensing capabilities and adhere to the motion model $\boldsymbol{x_{j, t}} = f(\boldsymbol{x_{j, t-1}}, \boldsymbol{u_{j, t-1}})$, which can be expressed as follows:
\begin{equation}
\left(\begin{array}{c}
x_{j,t}^1 \\
x_{j,t}^2 \\
\phi_{j,t}
\end{array}\right) =
\left(\begin{array}{c}
x_{j,t-1}^1 \\
x_{j,t-1}^2 \\
\phi_{j,t-1}
\end{array}\right) + \left(\begin{array}{c}
u_{j,t-1} \cos \left(\phi_{j,t-1}\right) \\
u_{j,t-1} \sin \left(\phi_{j,t-1}\right) \\
\omega_{j,t-1}
\end{array}\right).
\end{equation}

\subsubsection{State model of BMAV} 
\majorrevise{
The heterogeneous MAV swarm comprises $N$ ($N > 1$) BMAVs. 
The number of BMAVs is greater than or equal to the number of AMAVs ($N \geq M$).
}
Each BMAV $B_i$'s state at time $t$ includes both its location and motion command. The location of $B_i$ is denoted as $\boldsymbol{y_{i, t}} = (y_{i,t}^1, y_{i,t}^2)$, and its motion command is $\boldsymbol{v_{i, t}} = (v_{i,t}^1, v_{i,t}^2)$.
The $B_i$ follows double integrator dynamics with Gaussian noise, consistent with the model in \cite{purohit2013sugarmap}:
\begin{equation}
\begin{gathered}
\left(\begin{array}{l}
y_{i, t+\delta}^1 \\
y_{i, t+\delta}^2
\end{array}\right)=\left(\begin{array}{l}
y_{i, t}^1 \\
y_{i, t}^2
\end{array}\right)+\delta\left(\begin{array}{l}
v_{i, t}^1+n_{i, t}^1 \\
v_{i, t}^2+n_{i, t}^2
\end{array}\right),\\
\text{$n_{i, t}^1$ is drawn from $p(n^1)$,} \\
\text{$n_{i, t}^2$ is drawn from $p(n^2)$.}
\end{gathered}
\label{y}
\end{equation}
The motion noise terms $n_{i, t}^1$ and $n_{i, t}^2$ are drawn from motion noise models $p(n^1)$ and $p(n^2)$ respectively, with $\delta$ ($\delta > 1$) representing the time interval between two commands.
The distributions $p(n^1)$ and $p(n^2)$, empirically obtained from the testbed, are specified as normal distributions with mean $\mu = 0$ and variance $\sigma$, expressed as a percentage of $v_{i, t}^1$ and $v_{i, t}^2$, consistent with the approach outlined in \cite{chen2020h}.
The location estimation of $B_i$ is denoted as $\boldsymbol{\widehat{y}_{i, t}}$, and its covariance matrix is denoted as $\Sigma_{i, t}$. 
The noisy motion measurement from $B_i$'s Inertial Measurement Unit(IMU) is represented as $\boldsymbol{\dot{v}_{i, t}}$.

\subsubsection{Observation model of AMAV}
The observation angle of the visual sensor on AMAVs is denoted by $\phi$, and at time $t$, the Field of View (FoV) of $A_j$ is defined as follows:
\begin{equation}
\begin{aligned}
& F_{j, t}=\{(x, y) \mid \\
& \quad \quad 0<\sqrt{\left(y-x_{j,t}^2\right)^2+\left(x-x_{j,t}^1\right)^2}<r_m \quad \bigcap \\
& \quad -\frac{\phi}{2}<\arctan\left(\left(y-x_{j,t}^2\right)\left(x-x_{j,t}^1\right)\right)-\phi<\frac{\phi}{2}\}, 
\end{aligned}
\label{F_t}
\end{equation}
where $r_m$ represents the maximum observable distance constrained by the sensor's observation technique.

When $B_i$ falls within the FoV of $A_j$ at $t$, $A_j$ generates a noisy observation $z_{i, j, t}$ for $B_i$, comprising the range $r_{i, j, t}$ and bearing $\alpha_{i, j, t}$. Both $r_{i, j, t}$ and $\alpha_{i, j, t}$ are relative to $A_j$.
\begin{equation}
\begin{gathered}
\begin{aligned}
z_{i, j, t}
&=h(\boldsymbol{x_{j, t}}, \boldsymbol{y_{i, t}}) + \boldsymbol{n_{i, j, t}},\\
h(\boldsymbol{x_{j, t}}, \boldsymbol{y_{i, t}})
&:=\left[\begin{array}{c}
r_{i, j, t} \\
\alpha_{i, j, t}
\end{array}\right]\\
&:=\left[\begin{array}{c}
\sqrt{\left(y_{i, t}^1-x_{j, t}^1\right)^2+\left(y_{i, t}^2-x_{j, t}^2\right)^2} \\
\arctan \left(\left(y_{i, t}^2-x_{j, t}^2\right)\left(y_{i, t}^1-x_{j, t}^1\right)\right)-\phi
\end{array}\right]\\
\boldsymbol{n_{i, j, t}}
&:= [n_{i, j, t}^r, n_{i, j, t}^\alpha]^T
\end{aligned},\\
\text{$n_{i, j, t}^r$ is drawn from $p(n^r)$}, \\
\text{$n_{i, j, t}^\alpha$ is drawn from $p(n^\alpha)$}.
\end{gathered}
\label{z}
\end{equation}
The noise models for range and bearing measurements are represented by $p(n^r)$ and $p(n^\alpha)$, respectively, both empirically obtained from our testbed.
These models, $p(n^r)$ and $p(n^\alpha)$, are defined as normal distributions with mean $\mu = 0$ and variance $\sigma$, presented as a percentage of $r_{i, j, t}$ and $\alpha_{i, j, t}$.

\subsubsection{Linearization of Observation model} 
\majorrevise{
To linearize the observation model, we compute the Jacobian matrix of $h(\boldsymbol{x}, \boldsymbol{y})$ using Taylor expansion}, by calculating the gradient with respect to the location estimation of $\boldsymbol{y}$:
\begin{equation}
\setlength\abovedisplayskip{3pt}
\setlength\belowdisplayskip{3pt}
\begin{aligned}
&\nabla_{\boldsymbol{y}} h(\boldsymbol{x}, \boldsymbol{y})\\
&=\frac{1}{r(\boldsymbol{x}, \boldsymbol{y})}\left[\begin{array}{cc}
\left(y^1-x^1\right) & \left(y^2-x^2\right) \\
-\sin \left(\phi+\alpha(\boldsymbol{x}, \boldsymbol{y})\right) & \cos \left(\phi+\alpha(\boldsymbol{x}, \boldsymbol{y})\right)
\end{array}\right].
\end{aligned}
\label{linearize}
\end{equation}

To simplify notation, we let
$\boldsymbol{y_t} := {[\boldsymbol{y_{1, t}^T}, \dots, \boldsymbol{y_{N, t}^T}]}^T$, 
$\boldsymbol{x_t} := {[\boldsymbol{x_{1, t}^T}, \dots, \boldsymbol{x_{M, t}^T}]}^T$, 
$\boldsymbol{u_t} := {[\boldsymbol{u_{1, t}^T}, \dots, \boldsymbol{u_{M, t}^T}]}^T$, 
$\boldsymbol{\widehat{y}_t} := {[\boldsymbol{\widehat{y}_{1, t}^T}, \dots, \boldsymbol{\widehat{y}_{N, t}^T}]}^T$, 
${\Sigma_{t}} := {diag(\Sigma_{1, t}^T, \dots, \Sigma_{N, t}^T)}^T$. 
}


\begin{figure}[t]
    \centering    
        \includegraphics[width=1\columnwidth]{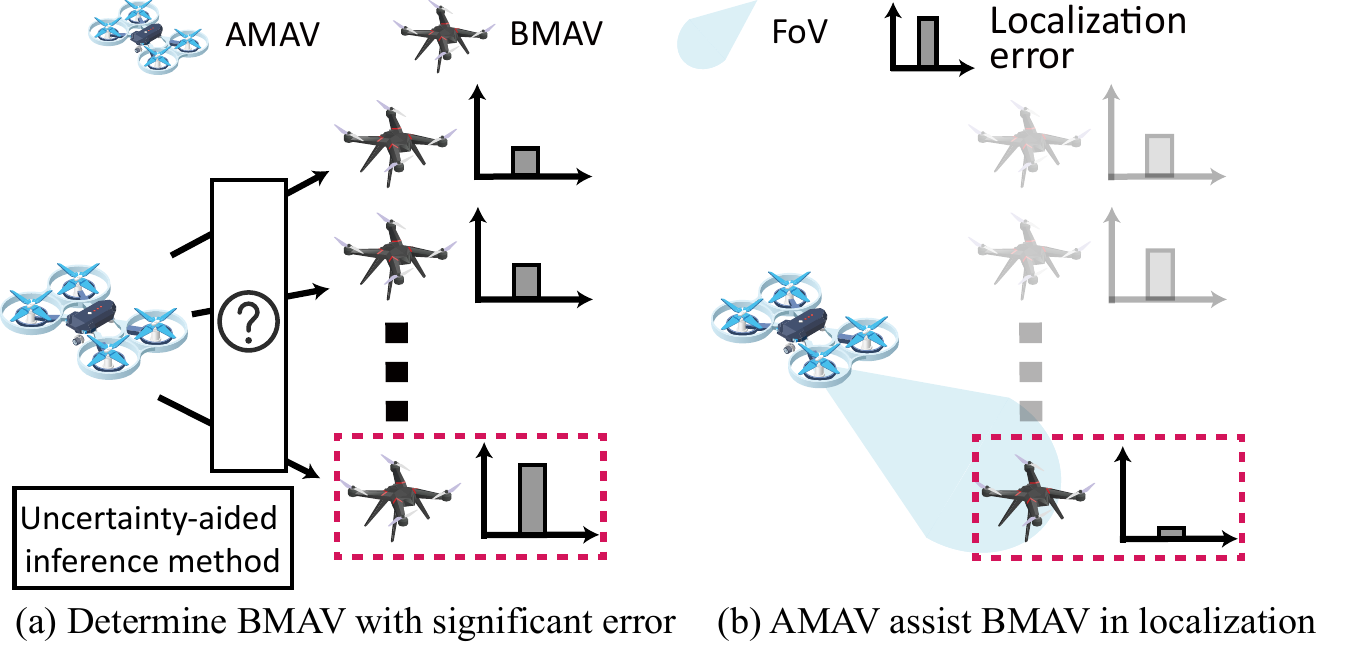}
        \vspace{-0.5cm}
    \caption{
    The error-aware joint pose estimation model operates as follows: (a) With the aid of the uncertainty-aided inference method, the AMAV identifies the BMAV with a higher error, and (b) subsequently generates observations for correction.}
    \label{error}
    \vspace{-0.5cm}
\end{figure} 

\subsection{Uncertainty-aided inference method} \label{3.2}
TransformLoc aims to efficiently allocate the sensing resources of AMAVs to generate observations for BMAVs, thereby minimizing localization errors. The localization error of a BMAV $B_i$ at time $t$ can be formally represented as:
\begin{equation}
\psi(i, t) := E\left[{||y_{i, t}-\widehat{y}_{i, t}||}^2\right].
\label{le}
\end{equation}
In this context, $y_{i, t}$ denotes the actual location of $B_i$, while $\widehat{y}_{i, t}$ represents the estimated location of $B_i$.
However, obtaining the actual locations of BMAVs is impractical due to their limited capabilities and the absence of external localization infrastructure.
Incorporating the Kalman filter-based model for estimating the location of BMAVs, we rely on a measure of estimation uncertainty to gauge the quality of BMAVs' location estimations. This approach eliminates the need to access the actual locations of BMAVs in order to evaluate localization quality.
To quantify the estimation uncertainty of BMAVs, we opt for the trace of the covariance matrix of the BMAV's state estimator \cite{ucinski2004optimal}. This is mathematically expressed as:
\begin{equation}
\psi(i, t) = tr\left(\Sigma_{i, t}\right),
\label{ob}
\end{equation}
where $\Sigma_{i, t}$ represents the trace of the covariance matrix of $B_i$ at time $t$. 
This indicator serves as a measure of estimation uncertainty, where a lower value signifies greater certainty. 
It facilitates AMAV to generate observations tailored to assist BMAVs with larger errors when dealing with BMAVs of varying errors, as depicted in \fig~\ref{error}a and \fig~ \ref{error}b.

\subsection{Joint estimation of BMAVs' location} \label{3.3}
The location estimation framework for BMAVs comprises two operations: (1) prediction based on the state model and noisy velocity measurements of BMAVs, which computes the prior distribution of BMAVs' location; (2) correction based on observations from AMAVs, which computes the posterior distribution of BMAVs' location. Further details regarding the joint estimation of BMAVs' location are elucidated in Algorithm \ref{estimation}.



\noindent $\bullet$ The \textit{Prediction of BMAVs from motion model} is outlined in Lines (1-3) of Algorithm \ref{estimation}. 
During this operation, the motion model, along with the noisy motion measurement $\dot{v}_{i, t}$ and the previous location estimation $\boldsymbol{\hat{y}_{i, t-1}}$ of BMAV $B_i$ at time $t$, are utilized to compute the prior distribution and the covariance matrix, denoted as $\boldsymbol{{y_{i, t}}^{-}}$ and ${\Sigma_{i, t}}^-$, respectively.


\noindent $\bullet$ \textit{Correction of BMAVs from observations} is depicted in Lines (4-11) of Algorithm \ref{estimation}.
During this operation, when BMAV $B_i$ falls within the FoV of AMAV $A_j$ at time $t$, the observation $z_{i, j, t}$ and the prior distribution of $B_i$'s location are utilized to compute the posterior distribution and the covariance matrix at time $t$, denoted as $\boldsymbol{{y_{i, t}}^{+}}$ and ${\Sigma_{i, t}}^+$, respectively.
The symbol $\eta$ in Line 8 represents a normalization constant.
If AMAV $A_j$ generates an observation for $B_i$, the location estimation $\boldsymbol{\widehat{y}_{i, t}}$ and the covariance matrix of estimation $\Sigma_{i, t}$ for $B_i$ at time $t$ are drawn from the posterior distribution $\boldsymbol{{y_{i, t}}^+}$ and ${\Sigma_{i, t}}^+$; otherwise, they are drawn from the prior distribution $\boldsymbol{{y_{i, t}}^-}$ and ${\Sigma_{i, t}}^-$.

\section{Similarity-instructed adaptive grouping-scheduling strategy} \label{5}


The BMAVs are dispersed across different locations and exhibit varying levels of localization errors. 
Due to the limited FoV and velocity of AMAV, it is crucial to identify the BMAVs it should allocate localization resources accordingly.
Furthermore, since both the AMAVs and BMAVs are continuously moving, the set of observable BMAVs for each AMAV is constantly changing. 
This dynamic nature necessitates the adaptation of resource allocation strategies, involving optimization in a high-dimensional decision space.

\revise{In this section, we present a \textit{Similarity-instructed adaptive grouping-scheduling} strategy for allocating sensing and computing resources of AMAVs to assist BMAVs in localization. 
We begin by outlining the mathematical formulation of the resource allocation problem (Sec. \ref{4.1}), followed by an introduction to the strategy. 
This strategy initially dynamically groups AMAVs and BMAVs based on spatial distance, leveraging the Voronoi diagram (Sec. \ref{4.2}). 
This step converts the complex many-to-many resource allocation problem into multiple one-to-many resource allocation subproblems. 
Subsequently, to plan the trajectory of each AMAV in a non-myopic manner and determine the optimal distance and angle, the strategy constructs search trees for each AMAV by involving several-step lookahead about BMAVs. 
Additionally, with the aim of further reducing search space, the strategy selectively prunes spatially close nodes with similar covariances related to BMAVs, thereby eliminating redundant search space (Sec. \ref{4.3}).
Finally, we delineate the planning method for BMAVs (Sec. \ref{4.4}).}

\subsection{Problem Formulation} \label{4.1}
\revise{
Our objective is to optimize the trace of the covariance of BMAVs’ location estimation $\Sigma_t$ to constrain the localization error within a finite time horizon $T$, taking into account the motion commands of each AMAV within this time span.
The mathematically formulated sensing resource allocation problem is as follows:
\begin{equation}
\min _{\boldsymbol{u_0}, \ldots, \boldsymbol{u_{T-1}}} \xi(T) = \sum_{t=1}^T\left(tr\left(\Sigma_t\right)\right),
\end{equation}
\text {\centerline{s.t.} }
\begin{align}
    & \boldsymbol{x_{t+1}} = f(\boldsymbol{x_{t}}, \boldsymbol{u_{t}}), t \in \{0,\dots, T-1\} \\
    & 0 \leq x_{j, t}^1, y_{i, t}^1 \leq L, \\
    & \quad j \in \{0,\dots, N\}, i \in \{0,\dots, M\}, t\in\{0,\dots, T-1\} \notag \\
    & 0 \leq x_{j, t}^2, y_{i, t}^2 \leq W, \\
    & \quad j \in \{0,\dots, N\}, i \in \{0,\dots, M\}, t\in\{0,\dots, T-1\} \notag \\
    & \Sigma_{t+1}=\rho_{t+1}^e\left(\rho_t^p\left(\Sigma_t\right), \boldsymbol{x}_{t+1}\right), t\in\{0,\dots, T-1\}.
\label{problem}
\end{align}
Eq. (9)  delineates the state transition equation of the AMAV, while Eq. (10) and (11) characterize the motion range of the AMAV. In Eq. (12), $\rho_t^e$ and $\rho_{t+1}^p$ denote the correction and prediction steps, respectively, elucidated in Sec. \ref{3.3}.}

\begin{algorithm}
	\renewcommand{\algorithmicrequire}{\textbf{Input:}}
	\renewcommand{\algorithmicensure}{\textbf{Output:}}
	\caption{$A_j$ assists $B_i$ for localization using noisy motion measurements and observations.} 
    	\begin{algorithmic}[1]
            \REQUIRE Prior estimate $\boldsymbol{\widehat{y}_{i, t-1}}$, covariance matrix $\Sigma_{i, t-1}$, noisy motion measurements $\dot{v}_{i, t-1}$, AMAV state $\boldsymbol{x_{j, t-1}}$, AMAV motion command $u_{j, t-1}$
            \ENSURE Updated estimate $\boldsymbol{\widehat{y}_{i, t}}$, updated covariance matrix $\Sigma_{i, t}$\\
                \textit{\textbf{\% Prediction of BMAV from motion model}}
    		\STATE Update prior distribution of $B_i$'s location at time $t$: $\boldsymbol{y_{i, t}}^{-} =\int p\left(\boldsymbol{y_{i,t}} \mid \boldsymbol{y_{i,t-1}}, \dot{v}_{i, t-1} \right) \boldsymbol{\hat{y}_{i,t-1}} d \boldsymbol{y_{i,t-1}}$;
          	\STATE Update covariance matrix ${\Sigma_{i, t-1}}^{-}$ from $\boldsymbol{y_{i, t}}^{-}$;
                \STATE Update $\boldsymbol{\widehat{y}_{i, t}}$ and $\Sigma_{i, t}$ from $\boldsymbol{y_{i,t}}^{-}$ and ${\Sigma_{i, t}}^{-}$;\\
                  \textit{\textbf{\% Correction of BMAV from observations}}
                \STATE Update the FoV of $A_j$, $F_{j, t}$, according to Eq. (\ref{F_t});
    		\IF{$B_i$ is in the FoV of $A_j$}
                \STATE Update observation $\boldsymbol{z_{i,j,t}}$ according to Eq. (\ref{z});
                \STATE Linearize the observation according to Eq. (\ref{linearize});
    		\STATE Update posterior distribution of $B_i$'s location at time $t$: $\boldsymbol{y_{i, t}}^{+}=\eta p\left(\boldsymbol{z_{i,j,t}} \mid \boldsymbol{y_{i,t}}\right) \boldsymbol{y_{i, t}}^{-}$;
                \STATE Update covariance matrix ${\Sigma_{i, t}}^{+}$ from $\boldsymbol{y_{i, t}}^{+}$;
    		\STATE Update $\boldsymbol{\widehat{y}_{i, t}}$ and $\Sigma_{i, t}$ from $\boldsymbol{y_{i,t}}^{+}$ and  ${\Sigma_{i, t}}^{+}$;
    		\ENDIF
	\end{algorithmic} 
\label{estimation}
\end{algorithm}

\subsection{Graph-based adaptive grouping} \label{4.2}
The BMAVs are located in various locations with varying localization errors. 
When AMAVs generate observations for BMAVs, a single AMAV's may waste sensing resources by traversing between different BMAVs. 

In this section, we propose a method to dynamically group BMAVs, enabling each AMAV to concentrate its sensing resources on a single group. 
However, grouping BMAVs and allocating them to different AMAVs poses a combinatorial optimization challenge, particularly with a large number of BMAVs and AMAVs, given its inherently NP-hard nature. 
TransformLoc initially partitions the entire area into non-overlapping regions based on the locations of AMAVs. 
Subsequently, each AMAV allocates its sensing resources to BMAVs within the nearest region for a duration of $\delta$, which represents the control command interval of BMAVs. 
Consequently, all BMAVs are divided into non-overlapping groups, each assigned to a different AMAV.

\begin{figure*}[t]
    \centering
        \includegraphics[width=2\columnwidth]{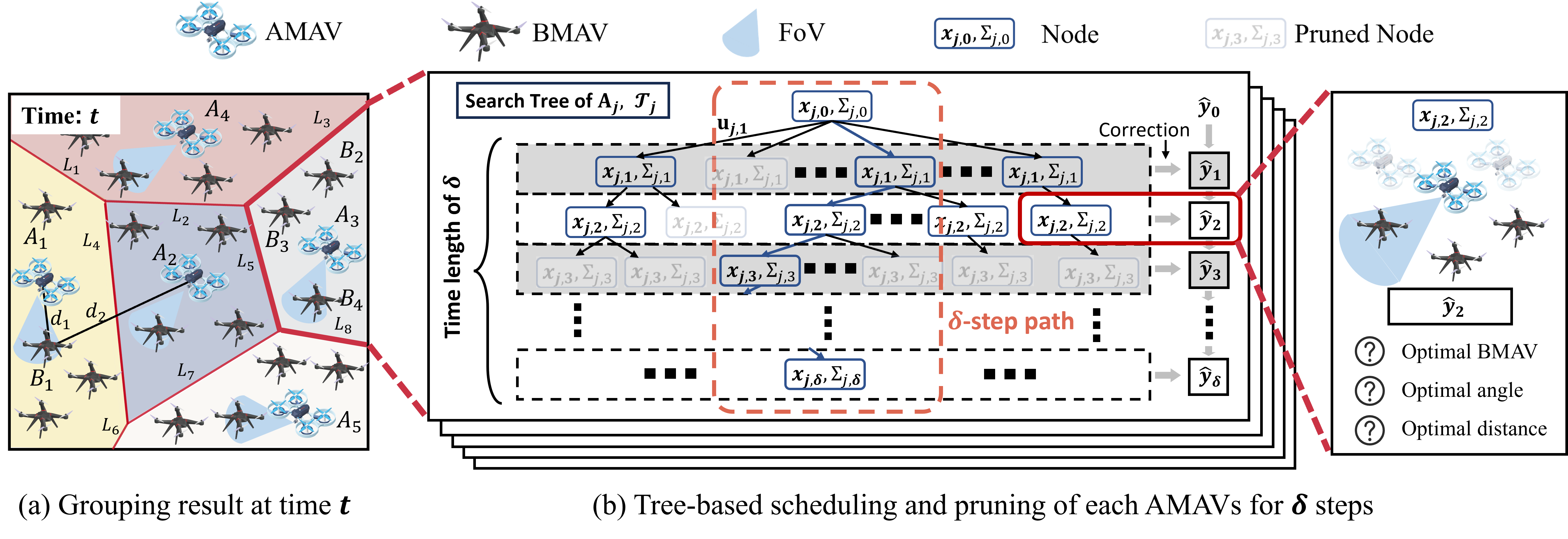}
    \vspace{-0.4cm}
    \caption{
    \revise{The similarity-instructed adaptive grouping-scheduling strategy aims to address the resource allocation challenge. 
    Initially, it leverages the Voronoi diagram to group MAVs, effectively decoupling the resource allocation problem. 
    Subsequently, the strategy constructs search trees for each AMAV, employing lookahead techniques regarding BMAVs, and employs a similarity-instructed method to prune redundant nodes. This results in $\delta$-step paths for AMAVs. 
    The scheduling of AMAVs is conducted in a non-myopic manner, enabling them to assist BMAVs with significant errors at optimal distances and angles. 
    Here, $x_{j, 0}$ represents the state of $A_j$, $\Sigma_{j, 0}$ signifies the covariance of assigned BMAVs' estimation maintained by $A_j$, and $\hat{y}_0$ denotes the estimation location of BMAVs.}
    }
    \label{grouping-scheduling}
    \vspace{-0.4cm}
\end{figure*} 

\subsubsection{Graph-based region partitioning} 

\majorrevise{
The partitioning method is grounded in the Voronoi diagram \cite{aurenhammer1991voronoi}. Under this approach, each AMAV is assigned a region containing points equidistant to or closer to it than to any other AMAVs. The steps followed to create the Voronoi regions include:  
\textit{(i) Seed Point Selection}, which identifies the positions of all AMAVs in the operational area. These positions serve as the seed points or generating sites for the Voronoi diagram.  
\textit{(ii) Distance Calculation}, where the Euclidean distance from each point in the space to every seed point is computed, determining the closest seed point for each location.  
\textit{(iii) Bisector Construction}, which calculates the perpendicular bisector of the line segment connecting each pair of seed points. These bisectors define the boundaries where points are equidistant to two seed points.  
\textit{(iv) Region Formation}, where all bisectors are combined to partition the space into non-overlapping regions. Each region is bounded by segments of bisectors and contains all points closest to its corresponding seed point.  
\textit{(v) Boundary Handling}, which constrains the Voronoi diagram to the predefined operational area for finite spaces, clipping any regions that extend beyond these boundaries to ensure practical applicability.  
This approach ensures an efficient and equitable partitioning of the operational area, facilitating effective task allocation among AMAVs.
Fig. \ref{grouping-scheduling}a illustrates this partitioning method with five AMAVs and several BMAVs. 
The outcome is depicted by region boundaries labeled as \textit{L1} - \textit{L8}, separated by perpendicular bisectors of neighboring AMAVs.
Any BMAV within region of $A_1$ is closer to it than to any other AMAV, indicating the distance between the BMAV and $A_1$ is shorter than its distance to any other AMAV (i.e., \textit{d1} $<$ \textit{d2}).
}

\subsubsection{Group and assignment of MAVs}
BMAVs located within the corresponding region of an AMAV are grouped, and the AMAV allocates its sensing resources to aid in their localization for a duration of $\delta$. 
For example, AMAV $A_3$ generates observations for BMAVs $B_2$, $B_3$ and $B_4$ within its region in Fig. \ref{grouping-scheduling}a. 
The group refreshes after an interval of $\delta$. 
It is important to note that the boundaries of a region for a given AMAV are determined solely by the locations of neighboring AMAVs.
\majorrevise{
These boundaries can be computed by identifying the perpendicular bisectors between pairs of adjacent AMAVs. 
}
If no BMAVs are present within an AMAV's region, it allocates sensing resources to all BMAVs over the duration of $\delta$.

\subsection{Search tree-based non-myopic scheduling} \label{4.3}
\revise{
The AMAV encounters two primary challenges when generating observations for its assigned BMAVs:
$(i)$ Observation noise is proportional to the distance between the AMAV and the BMAV (Eq. (\ref{z})). Closer proximity reduces noise, but the number of BMAVs that can be assisted by an AMAV is constrained by both the size of the BMAV and the AMAV's FoV (Eq. (\ref{F_t})). Balancing this trade-off is essential for precise localization of BMAVs.
$(ii)$ BMAVs move to destinations using Eq. (\ref{y}) while the AMAVs generate observations, necessitating the AMAVs to adjust for displacement. This calls for non-myopic AMAV planning, which anticipate movements of BMAVs several steps ahead. 
However, the exponential increase in the AMAV's action space with the planning horizon results in high computational complexity.

In this section, we introduce the incorporation of a search tree-based scheduling strategy for the non-myopic resource allocation of AMAVs. 
The key concepts are as follows:
$(i)$ BMAVs receive commands at time intervals of $\delta$, enabling AMAV to acquire BMAV motion commands within $\delta$. These motion commands are utilized to schedule paths of AMAVs with a $\delta$-step lookahead for BMAVs.
$(ii)$ To address the high computational complexity, the search tree is pruned. Specifically, if two nodes in the search tree have similar AMAV locations, the node with the higher BMAV uncertainty may be pruned from the search tree.

\subsubsection{Search tree construction}

We construct search trees for each AMAV.
As depicted in Fig. \ref{grouping-scheduling}b, we illustrate an example of constructing a search tree $\mathcal{T}_j$ for $A_j$. 
$\mathcal{T}_j$ encompasses potential trajectories that $A_j$ can pursue, originating from an initial location and covariance pair $(\boldsymbol{x_{j,0}}, \Sigma_{j,0})$.
Here, $\boldsymbol{x_{j,0}}$ denotes the starting location of $A_j$, and $\Sigma_{j,0}$ represents the initial covariance of the location estimations of the BMAVs assigned to $A_j$. 
The nodes within the search tree at level $t \leq \delta$ correspond to reachable locations for $A_j$ and are denoted as $(\boldsymbol{x_{j,t}}, \Sigma_{j,t})$. 
At each location, the AMAV assesses the distance and angle to observable BMAVs to determine the optimal distance and angle for generating observations for BMAVs with significant errors.

We discretize the control space of the AMAV, where $A_j$ possesses a finite set of control options $\mathcal{U}$. 
For each option $\boldsymbol{u_{j, t}}$, there exists an edge originating at node $(\boldsymbol{x_{j,t}}, \Sigma_{j,t})$ and leading to node $(\boldsymbol{x_{j,t+1}}, \Sigma_{j,t+1})$, determined by evaluating the state model of the AMAV. 
The estimation location of the corresponding BMAVs, $\boldsymbol{\hat{y}_t}$, is computed by evaluating the motion model of the BMAV, the observation model between the BMAV and AMAV, and the Algorithm \ref{estimation}.

\subsubsection{Similarity-instructed Search tree pruning}


The final level of $\mathcal{T}_j$ encompasses $\mathcal{O}\left(\mathcal{U}^\delta\right)$ nodes, rendering it impractical to compute in real-time for extended horizons $\delta$ by the AMAV.
To tackle this challenge, we employ pruning strategies to reduce the computational burden.
During the construction of $\mathcal{T}_j$, if two nodes exhibit identical physical locations at a particular level $t$ but differ in the estimation covariance of the assigned BMAVs' location ($\Sigma_{j, t}$), one of the nodes is deemed less informative and is thus considered for deletion.
This pruning process is carried out according to the following mathematical criteria:

\noindent $\bullet$ \textit{Definition 1 ($\epsilon$-Algebraic Redundancy) \cite{vitus2012efficient}}:
Given $\epsilon \geq 0$ and a finite set $\left\{\Sigma_i\right\}_{i=1}^K \subset S$, a matrix $\Sigma \subset S$ is $\epsilon$-Algebraic Redundancy with respect to $\left\{\Sigma_i\right\}_{i=1}^K$ if there exist non-negative constants $\{\alpha_i\}_{i=1}^K$ such that 
\begin{equation}
\sum_{i=1}^K \alpha_i=1 
\quad \text { and } \quad 
\Sigma+\epsilon I_n \succeq \sum_{i=1}^K \alpha_i \Sigma_i,
\end{equation}
where $I_n$ denotes the identity matrix. 
A node in the search tree can be pruned if it exhibits $\epsilon$-Algebraic Redundancy with other nodes on the same level, as it provides less information compared to the remaining nodes.

\noindent $\bullet$ \textit{Definition 2 (Trajectory $\sigma$-Crossing) \cite{atanasov2014information}}: 
Two trajectories $\pi^1$ and $\pi^2$ are said to $\sigma$-cross at time $t \in [1,\delta]$ if their distance $d_{\chi}(\pi^1, \pi^2)$ is less than or equal to $\sigma$ for $\sigma \geq 0$, where $d_{\chi}$ measures the distance between trajectories. 
If a node in the search tree $\mathcal{T}_j$ is $\sigma$-crossing with other nodes at the same level, it implies that these nodes are located in close proximity to each other.


We apply $\epsilon$-Algebraic Redundancy and Trajectory $\sigma$-Crossing criteria to determine which nodes at level $t$ of the search tree can be pruned.
Increasing $\epsilon$ leads to fewer nodes being retained in each level of the tree, approaching a greedy policy, while decreasing $\epsilon$ preserves more nodes.
As illustrated in Fig. \ref{grouping-scheduling}b, we demonstrate the construction and pruning of level $3$ of AMAV $A_j$'s search tree to exemplify the application of these criteria.
Initially, we retain the node with the minimum $tr(\Sigma_{j, 3})$.
Subsequently, we assess other nodes at level $3$.
If a node does not $\sigma$-cross with the nodes we have reserved, or if it does $\sigma$-cross but its estimation covariance of BMAVs' location is not $\epsilon$-redundant with the reserved nodes, we reserve this node. 
Otherwise, we prune this node.
This process eliminates less informative nodes that are spatially close.
Finally, we select node with minimum $tr(\Sigma_{j, \delta})$ at level $\delta$ and schedule $A_j$ for a period of $\delta$.}

\subsection{Scheduling of BMAV} \label{4.4}



\begin{figure*}[t]
    \centering
        \setlength{\abovecaptionskip}{0.cm}
        \includegraphics[width=2\columnwidth]{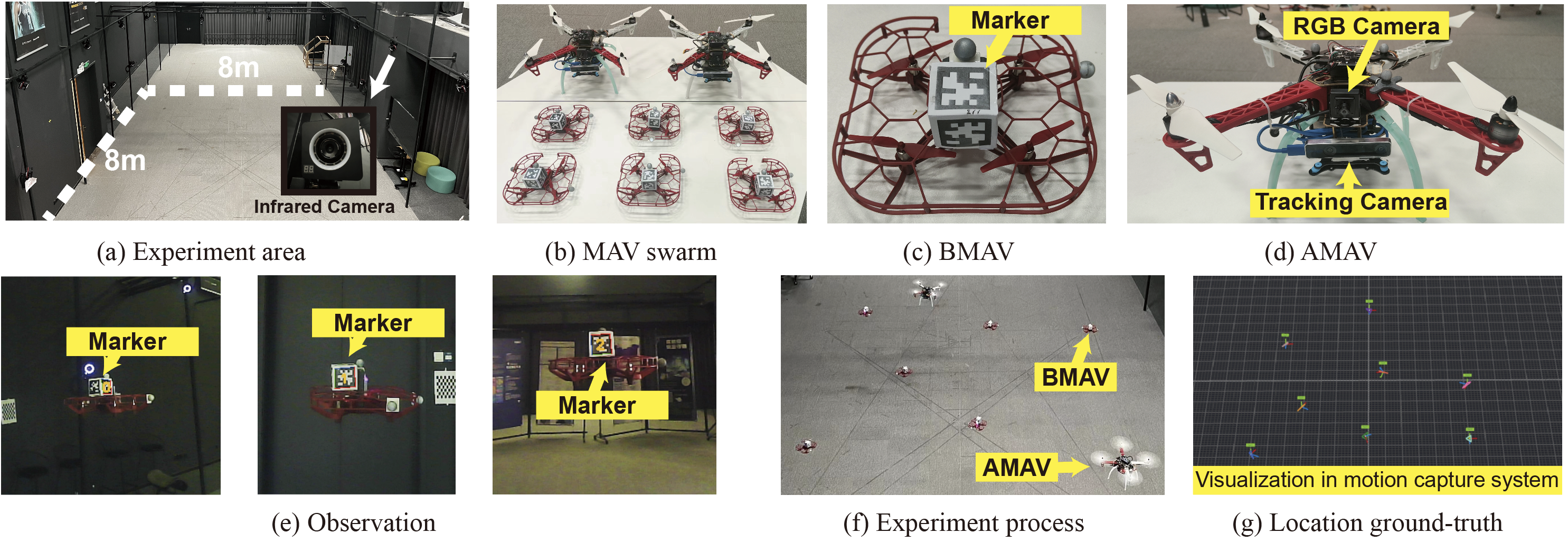}
    \caption{
    \revise{Experimental Setup and Implementation of Heterogeneous MAV Swarm. AMAVs utilize tracking cameras for localization and visual sensors for observation generation. BMAVs are equipped with AprilTag markers for recognition and localization. Ground truth data is obtained from a motion capture system. The experiment area measures $8m \times 8m$.}}
    \label{setup}
    \vspace{-0.6cm}
\end{figure*} 

The BMAVs' locations are initialized before takeoff, providing accurate initial localization. 
However, their sensing capabilities are limited, relying solely on the dead-reckoning algorithm, which integrates data from IMUs and optical flow sensors. 
Over time, the accuracy of their localization decreases due to reliance on this algorithm, leading to increased localization errors. 
Additionally, BMAVs have constrained computing capabilities, making advanced methods like SLAM using heavy sensors impractical.
Although AMAVs can offer observations for BMAVs, they cannot provide real-time assistance to all BMAVs due to the outnumbering of BMAVs by AMAVs. 

To tackle these challenges, TransformLoc utilizes location estimation for BMAV navigation. A lightweight planning algorithm based on artificial potential fields is employed, ensuring collision avoidance. 
TransformLoc maintains BMAVs' location estimates and generates motion commands at intervals of $\delta$ based on their distance from the destination and the obstacles. 
BMAVs navigate within a force field, where the destination exerts an attractive force proportional to the distance, while obstacles generate repulsive forces proportional to the distance. 
This approach allows BMAVs to dynamically adjust their motion as they approach the destination, reducing velocity and improving navigation success.

\section{Evaluation}\label{6}
\subsection{Implementation and Methodology}

\subsubsection{Testbed Implementation}
As illustrated in Fig. \ref{setup}, we implement TransformLoc using DJI Robomaster TTs (BMAVs) and industry drones (AMAVs) built on Pixhawk, one of the most widely used autopilot systems, to validate its performance in the real world. 
The AMAV is equipped with an Intel(R) T265 tracking camera for localization and an RGB camera with a FoV of 120 degrees for observation generation. 
Each BMAV is equipped with an IMU and a downward-facing optical flow sensor. 
Additionally, each BMAV mounts a $3cm \times 3cm$ AprilTag for recognition and observation generation \cite{wang2016apriltag} (Fig. \ref{setup}b). 
The AMAV utilizes ArduPilot frameworks for motion control (Fig. \ref{setup}c). 
Ground truth data with millimeter-level accuracy is provided by a motion capture system operating at 240 FPS in the experiment area measuring $8m \times 8m$ (Fig. \ref{setup}a and Fig. \ref{setup}d). 
We assess the robustness of TransformLoc using a physical-feature-based simulator in an experiment area similar to that shown in Fig. \ref{setup}(a).

\begin{figure*}[t]
\setlength{\abovecaptionskip}{0.cm} 
\setlength{\belowcaptionskip}{-0.4cm} 
\setlength{\subfigcapskip}{-0.1cm}  
\centering
    \subfigure[The CDF of ATE]{
        \centering
            \includegraphics[width=0.65\columnwidth]{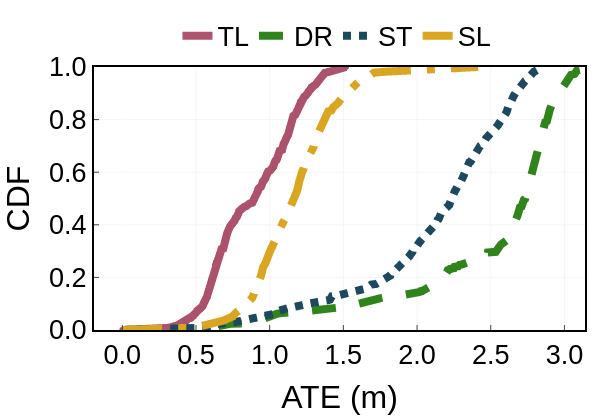}
    }
    \subfigure[Success rate within 200s]{
        \centering
            \includegraphics[width=0.65\columnwidth]{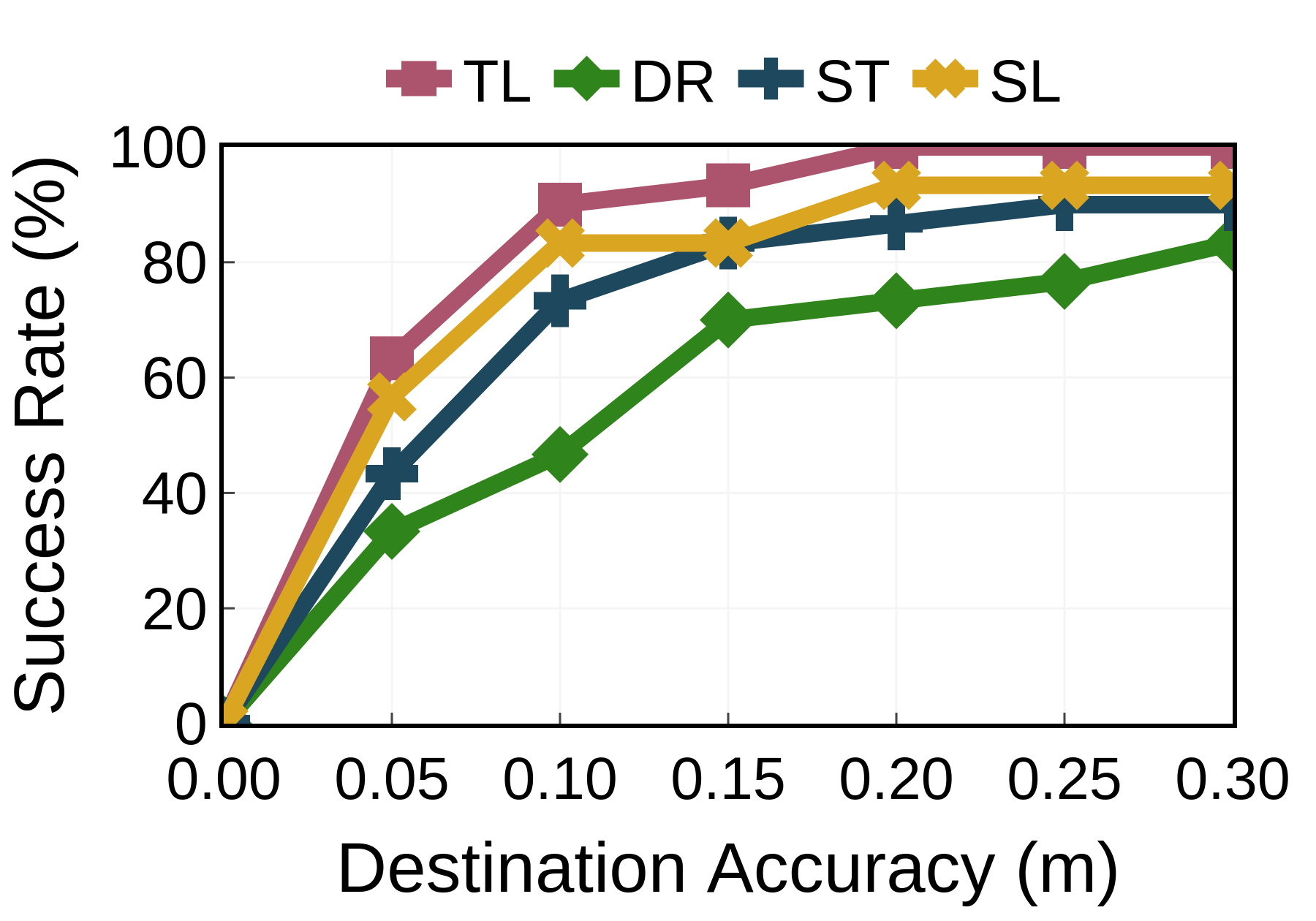}
    }
    \subfigure[Success rate with 0.2m accuracy]{
        \centering
            \includegraphics[width=0.65\columnwidth]{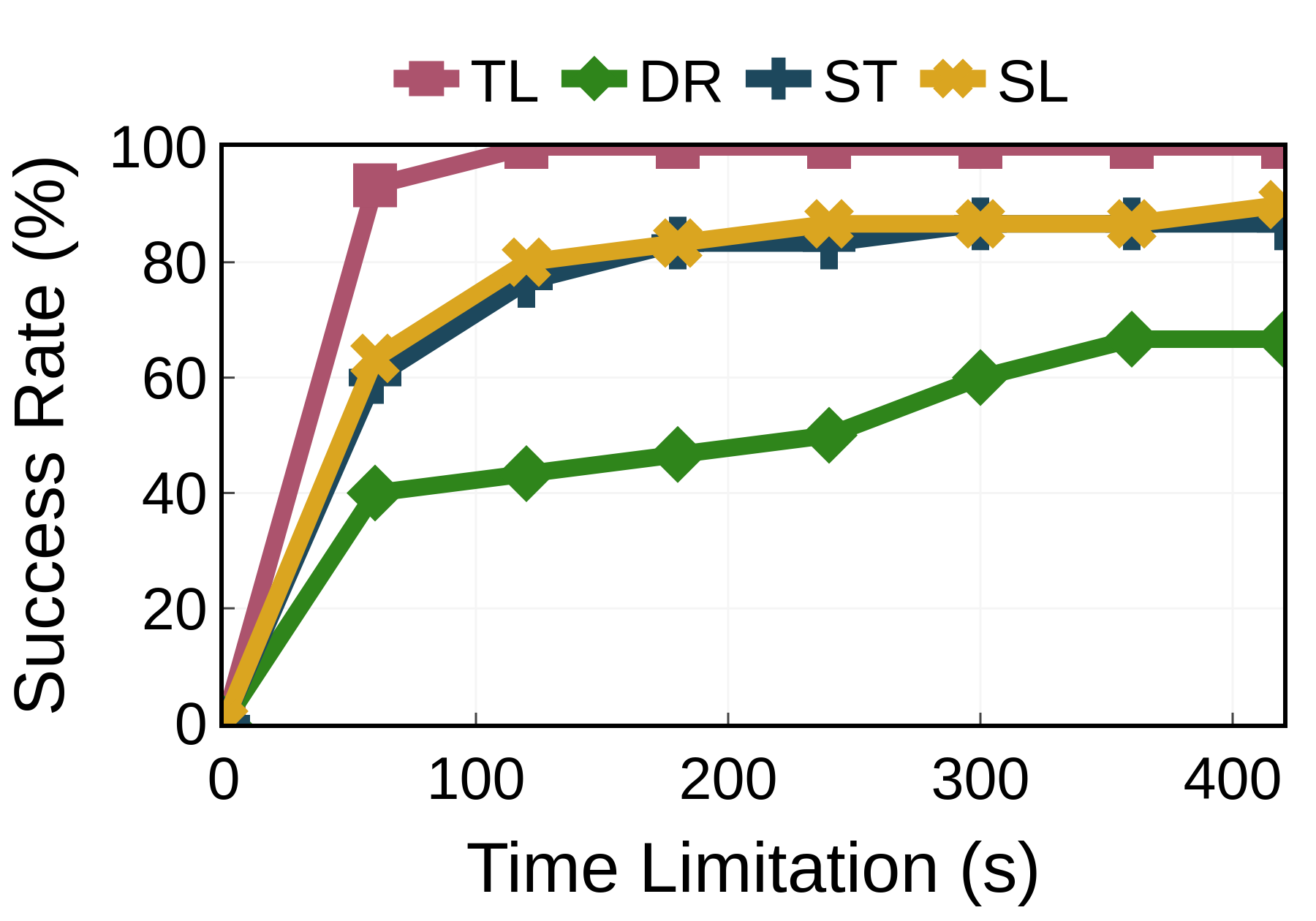}
    }
\caption{Overall performance in in-field experiments with two AMAVs and six BMAVs. 
}
\label{in-field}
\vspace{-0.4cm}
\end{figure*}

\begin{figure*}[t]
\setlength{\abovecaptionskip}{0.cm} 
\setlength{\belowcaptionskip}{-0.4cm} 
\setlength{\subfigcapskip}{-0.1cm}  
\centering
    \subfigure[The CDF of ATE]{
        \centering
            \includegraphics[width=0.65\columnwidth]{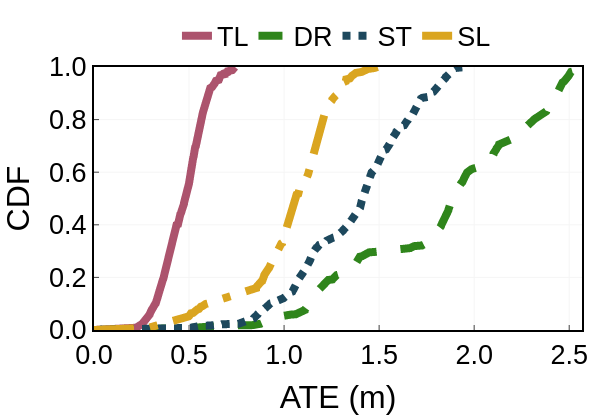}
    }
    \subfigure[Success rate within 200s]{
        \centering
            \includegraphics[width=0.65\columnwidth]{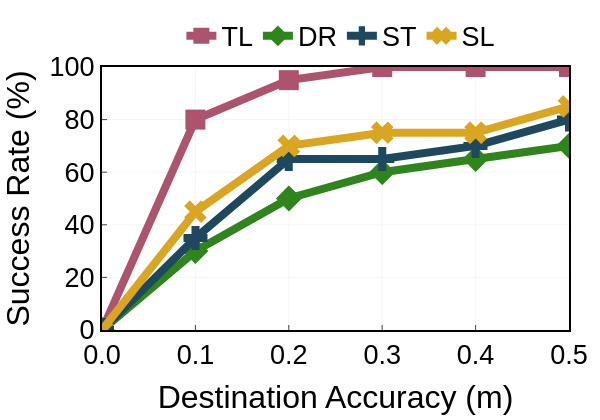}
    }
    \subfigure[Success rate with 0.2m accuracy]{
        \centering
            \includegraphics[width=0.65\columnwidth]{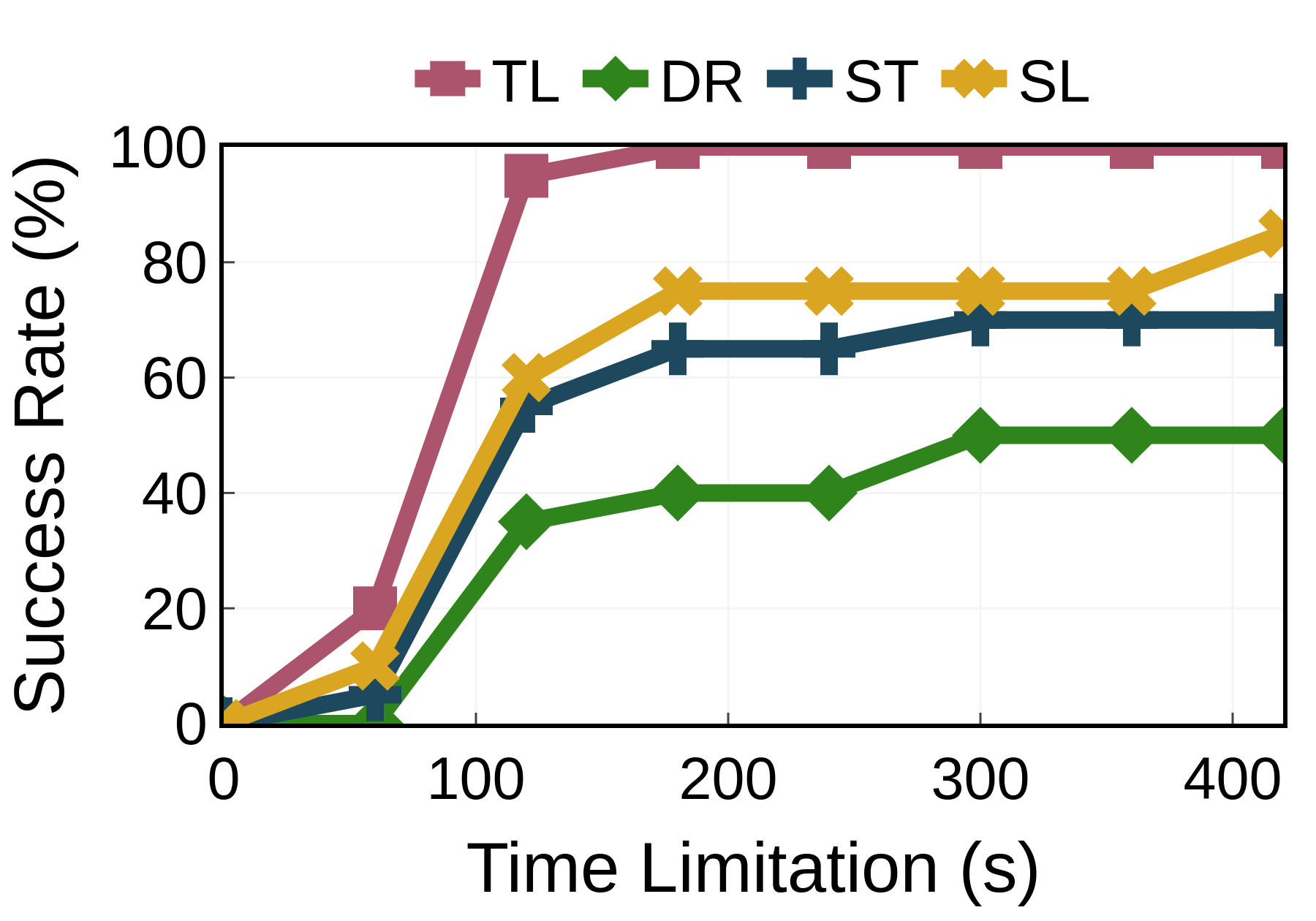}
    }

\caption{Overall performance in physical feature-based simulation with five AMAVs and twenty BMAVs. 
}
\label{simOverall}
\vspace{-0.4cm}
\end{figure*}

    
    
    
    
    

    
    

\subsubsection{Experiment setting}
The BMAVs are capable of reaching a maximum velocity of 0.5 m/s and operate with a command interval $\delta$ set to 5 steps. 
AMAVs possess an observation angle $\theta$ of 120 degrees and a maximum observation distance $r_m$ of 1 m. 
Control commands for AMAVs utilize motion primitives $\{(u, \omega)\}$ where $u \in \{0, 1, 3\}$ m/s and $\omega \in \{0, \pm1, \pm3\}$ rad/s. 
Performance evaluation was conducted through in-field experiments involving two AMAVs and six BMAVs, alongside simulations featuring five AMAVs and twenty BMAVs. 
During experiments, BMAVs navigated towards the perimeter of the room for environmental sensing. 
Simulations are conducted over a duration of 420 seconds, corresponding to the typical battery life of the DJI Robomaster TT drones used in the in-field experiments.  
The parameters for pruning the search tree, $\epsilon$ and $\sigma$, are set to 1 and 10, respectively. 
Noise models for BMAVs' motion and AMAVs' range and bearing measurements are determined empirically during in-field experiments and set to 20\%, 10\%, and 5\% of the measured values, respectively.

\subsubsection{Comparative Methods} 
We conduct a comprehensive evaluation of \textbf{TransformLoc (TL)} by comparing it against three baseline methods that operate without the need for a dedicated localization infrastructure. 
\textbf{(1) Dead-Reckoning (DR)} relies solely on measurements from motion sensors to estimate the locations of BMAVs \cite{borenstein1996navigating}. 
\textbf{(2) Station (ST)} employs stationary AMAVs positioned at fixed locations to provide observations for BMAVs \cite{xu2020learning}.
\textbf{(3) H-SwarmLoc (SL)} represents a state-of-the-art method, where a single AMAV is guided by reinforcement learning to support BMAVs in localization. To ensure a fair comparison, we modified SL to group MAVs in a predefined manner and direct AMAVs to generate observations \cite{wang2022h}.
\majorrevise{Additionally, we assess the latency performance of TransformLoc against \textbf{(4) CCM-SLAM \cite{schmuck2019ccm}}, a homogeneous collaborative SLAM method leveraging monocular cameras. 
The CCM-SLAM represents the first instance of a collaborative SLAM system that demonstrates bidirectional communication between agents and a central server during real-world flights involving three UAVs.
The CCM-SLAM achieves swarm localization through multi-UAV collaborative mapping, whereas TransformLoc realizes swarm localization by transforming the sensing and computational resources of AMAVs to BMAVs via observation. 
Thus, comparing TransformLoc with CCM-SLAM highlights the advantages of TransformLoc’s resource transformation method for swarm localization in terms of latency, because it avoids collaborative mapping with high computational overhead.}

\subsubsection{Evaluation Metrics}

The primary objective of TransformLoc is to enhance both the localization accuracy and navigation success rate of BMAVs. To assess its effectiveness, we employ two key performance metrics:
\textit{(1) Localization Error:} This metric evaluates the localization accuracy of BMAVs by comparing their estimated locations against ground truth data at each timestep. We quantify the localization error using the absolute trajectory error (ATE).
\textit{(2) Success Rate:} The success rate metric measures the proportion of BMAVs that successfully reach their destinations within predefined time and destination constraints.

\subsection{Overall Performance} 
\subsubsection{In-field experiments}

\fig~\ref{in-field} provides a detailed analysis of the outcomes obtained from in-field experiments.
To evaluate localization accuracy, we conduct a 420-second random walk for BMAVs, plotting the cumulative distribution function (CDF) of the ATE. TransformLoc achieves an ATE below $1.5m$, outperforming SL, ST, and DR, which achieve ATEs below $2.4m$, $2.8m$, and $3m$, respectively (as depicted in Fig. \ref{in-field}a).

Regarding navigation performance, we assessed the success rate of BMAVs over a 200-second duration under varying destination accuracy constraints (as illustrated in Fig. \ref{in-field}b). TransformLoc outperforms all baselines, achieving a remarkable 63\% success rate with a stringent destination accuracy constraint of 0.05m. Furthermore, under looser destination accuracy limits (ranging from 0.2m to 0.3m), TransformLoc attains a 100\% navigation success rate, surpassing the performance of the baselines.

Additionally, as the time limitation increases, the success rate of BMAVs with a destination accuracy constraint of 0.2m improves for all methods, benefiting from more time to reach the target (as shown in Fig. \ref{in-field}c). 
Notably, TransformLoc exhibits superior performance compared to the baselines, achieving a navigation success rate 22.3\% higher than SL, 25.6\% higher than ST, and 55.6\% higher than DR within strict time constraints (ranging from 60 to 180 seconds).
These results underscore the efficacy of TransformLoc in proficiently allocating AMAVs' resources to enhance the localization and navigation capabilities of BMAVs.
\begin{figure*}[t]
\setlength{\abovecaptionskip}{0.cm} 
\setlength{\belowcaptionskip}{-0.4cm} 
\setlength{\subfigcapskip}{-0.1cm}  
\centering
    \subfigure[Number of AMAVs]{
        \centering
        \includegraphics[width=0.46\columnwidth]{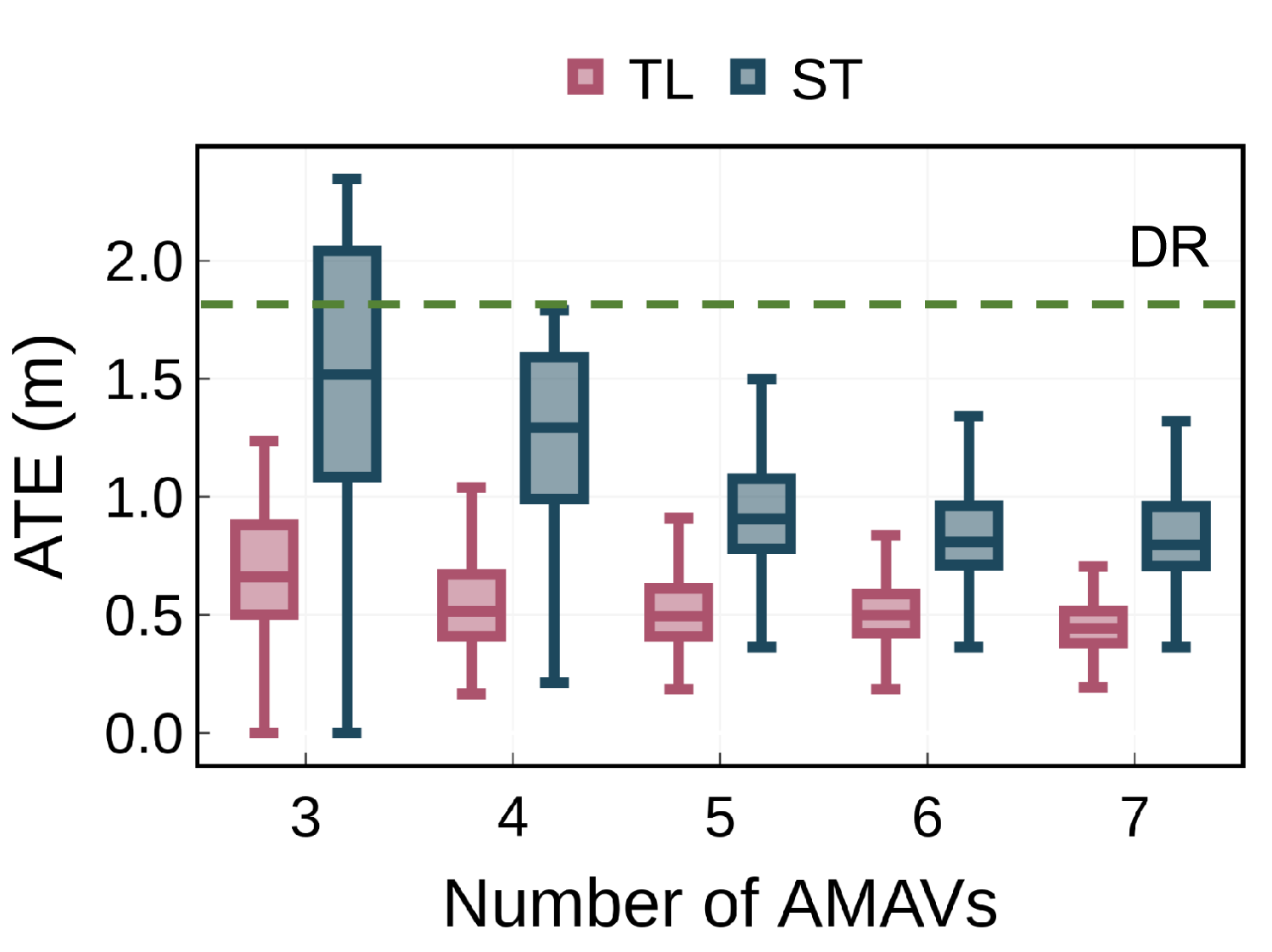}
    }
    \subfigure[Number of BMAVs]{
        \centering
            \includegraphics[width=0.50\columnwidth]{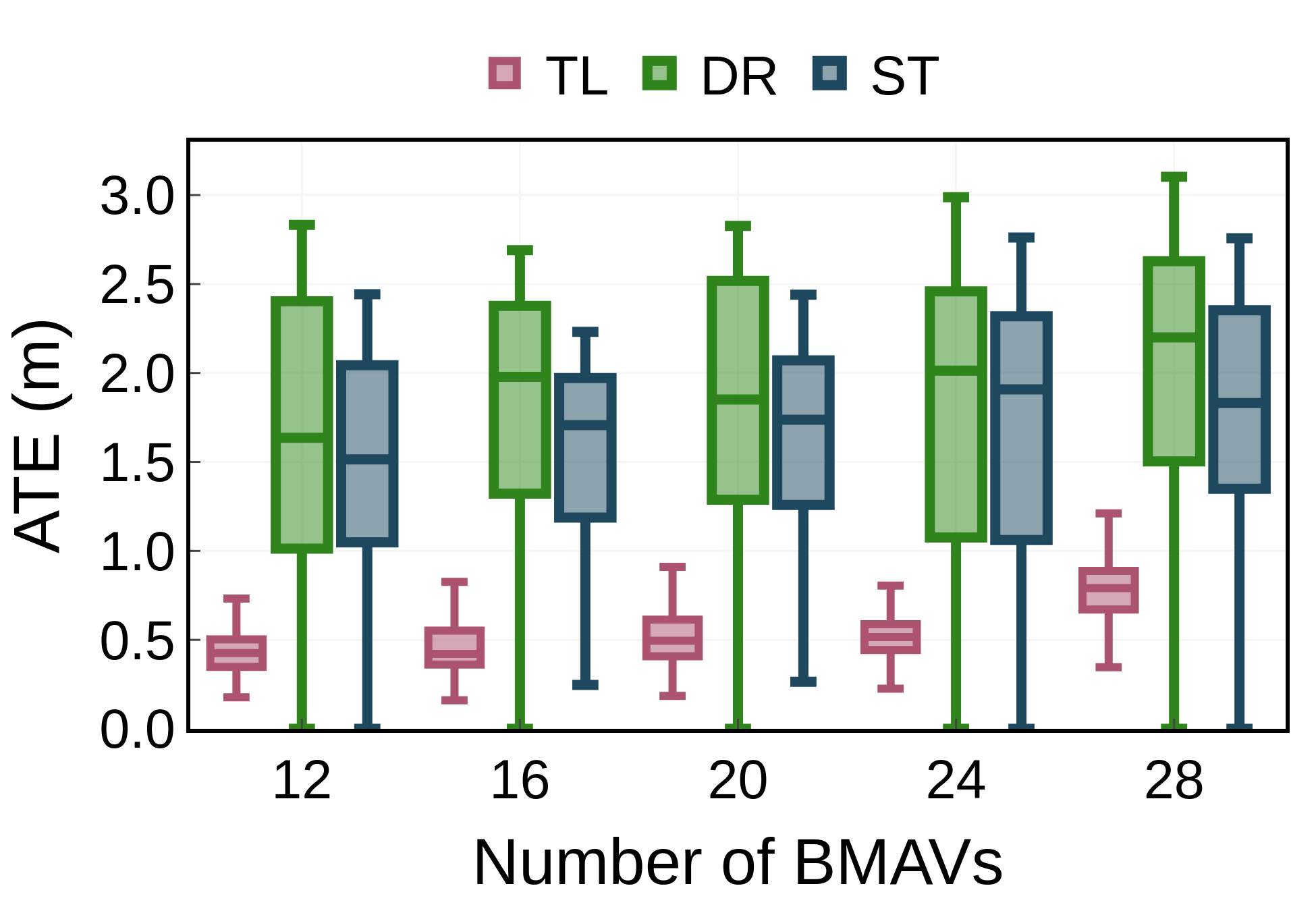}
    }
    \subfigure[\revise{Motion measurement noise}]{
        \centering
            \includegraphics[width=0.50\columnwidth]{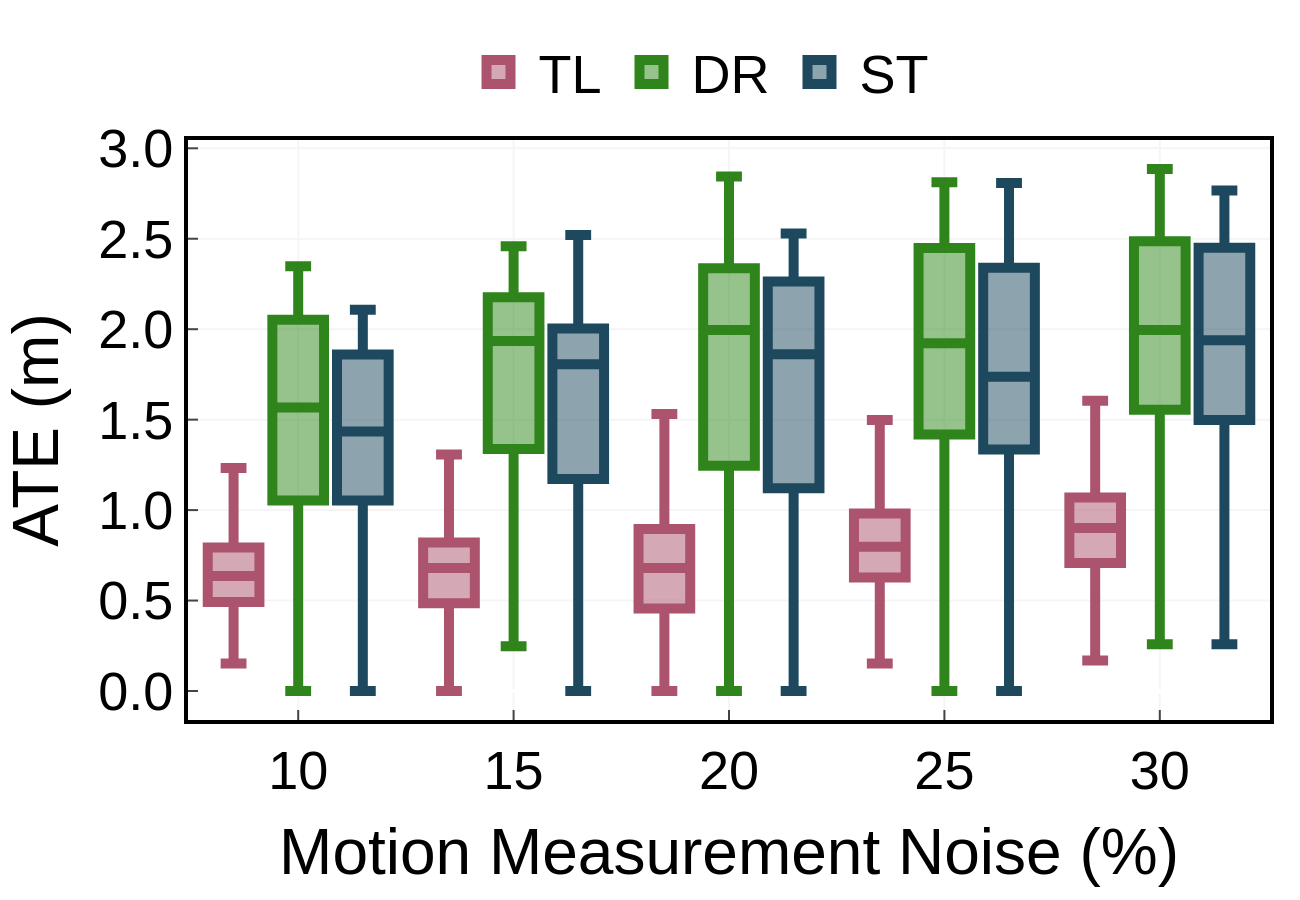}
    }
    \subfigure[\revise{Observation noise}]{
        \centering
            \includegraphics[width=0.46\columnwidth]{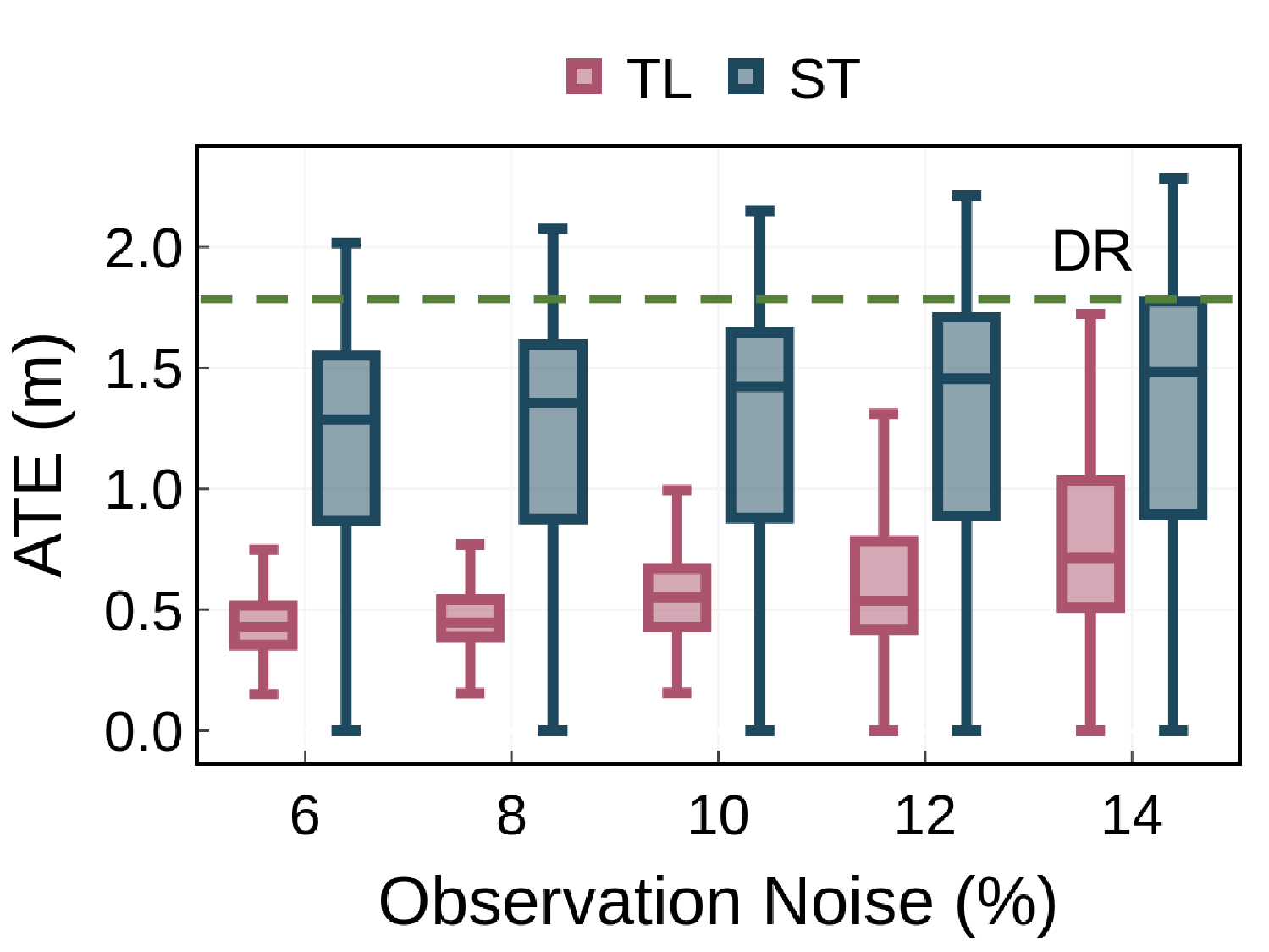}
    }
\caption{System Robustness Evaluation with five AMAVs and twenty BMAVs.}
\label{simrobust}
\vspace{-0.4cm}
\end{figure*}

\subsubsection{Physical feature-based simulations experiments}
\revise{
The simulations in this section evaluate the navigation success rate of the BMAVs under different destination constraints and time constraints. In the meanwhile, the localization error of all BMAVs is evaluated. 
To further verify TransformLoc's localization performance, we initiate a 420-second random walk for each BMAV. Fig. \ref{simOverall} (a) shows the CDF of localization errors.
TransformLoc achieves an impressive ATE below $0.7m$, while SL, ST, and DR maintain ATE below $1.5m$, $2m$, and $2.5m$ respectively.


We evaluate TransformLoc's navigation performance across various destination accuracy constraints by analyzing the success rate of BMAVs' navigation within a 200-second duration, as depicted in Fig. \ref{simOverall} (b). The success rate of BMAVs' navigation increases with increased destination accuracy, as the increase in destination accuracy means the farther away the destination may is when the BMAV is successfully navigated. 
TransformLoc consistently outperforms all baselines across different destination accuracy conditions. 
For strict destination accuracy (0.2m), TransformLoc achieves an impressive 80\% success rate, while all baselines fall below 50\%. Under less constrained destination accuracy (ranging from 0.4 to 0.6 m), TransformLoc maintains a consistent 30\% higher success rate than the baselines. At the loosest destination accuracy (0.8m), TransformLoc achieves a flawless 100\% navigation success rate, while the baselines remain below 80\%.
On average, TransformLoc exhibits substantial performance advantages over SL, ST, and DR, with margins of 25\%, 26.67\%, and 33.3\%, respectively. 
These findings underscore TransformLoc's effectiveness in accommodating varying destination accuracy levels and achieving consistently high navigation success rates.


To validate TransformLoc's navigation performance under various time constraints, we analyze the navigation success rate of BMAVs with a destination accuracy constraint of 0.2m, as depicted in Fig. \ref{simOverall} (c).
As the time limitation increases, all methods exhibit improved navigation success rates for BMAVs, given the extended duration to reach their destinations. TransformLoc consistently outperforms the other baselines across different time constraints.
Under strict time constraints (60-180 seconds), TransformLoc achieves a navigation success rate nearly 10\% higher than ST and SL. 
As time constraints become more relaxed (240-420 seconds), TransformLoc achieves a perfect 100\% navigation success rate, surpassing all other baselines.
On average, TransformLoc demonstrates significant performance advantages over SL, ST, and DR by 22.8\%, 25.6\%, and 42.5\%, respectively. 
By strategically dispatching AMAVs to provide observations for BMAVs near their final destination, TransformLoc effectively reduces localization errors and enhances the navigation success rate.
In summary, TransformLoc demonstrates the capability to efficiently allocate AMAVs' sensing resources, thereby improving BMAVs' navigation success rates across varying time constraints.
In conclusion, TransformLoc dispatches AMAVs to provide timely observations for BMAVs, resulting in lower localization errors.
}

\subsection{System Robustness Evaluation}

\subsubsection{Impact of Number of AMAVs}
\fig\ref{simrobust}a presents a comparison of the ATE of BMAVs across different methods and varying numbers of AMAVs. TransformLoc consistently achieves lower errors compared to DR, indicating that the inclusion of AMAVs contributes to the reduction in ATE.
As the number of AMAVs increases from 3 to 7, TransformLoc demonstrates a notable decrease in ATE to below $0.8m$. This reduction is substantial compared to ST, which only provides assistance to BMAVs when traversing over fixed AMAVs.
These findings underscore the significant impact of increasing the number of AMAVs in decreasing the ATE of BMAVs. 
Overall, the results highlight the effectiveness of TransformLoc in reducing localization errors through the strategic allocation of AMAVs.

\subsubsection{Impact of Number of BMAVs}


\fig\ref{simrobust}b depicts the performance of different methods in localizing varying numbers of BMAVs. 
As the number of BMAVs increases from 12 to 28, the ATE for all methods also increases. This trend is attributed to a smaller percentage of BMAVs being assisted by AMAVs as the overall BMAV count rises.
However, even with the increasing number of BMAVs, TransformLoc consistently achieves an ATE of less than $1.2 m$. This result is more than 57\% lower than any other baseline method. 
These experimental findings highlight the efficacy of TransformLoc in efficiently distributing the sensing resources of AMAVs to cater to different numbers of BMAVs, thereby ensuring a consistently low localization error.

\revise{
\subsubsection{Impact of Motion measurement noise of BMAV}

Motion measurement noise plays a pivotal role in determining the accuracy of BMAV localization. 
As this noise level increases, so does the localization error of BMAVs. 
The performance of TransformLoc under varying levels of motion measurement noise is assessed by plotting the ATE in Fig. \ref{simrobust}c. 
Even amidst increasing noise levels, TransformLoc consistently achieves a lower ATE compared to other baseline methods. 
Notably, TransformLoc maintains an error of less than $1.6m$ even when motion measurement noise reaches 30\%. 
These results underscore TransformLoc's ability to effectively allocate AMAVs' sensing resources for timely observation generation, particularly in scenarios with significant motion measurement noise.

\subsubsection{Impact of observation noise of AMAV}


Experimental results from the testbed highlight the substantial influence of range measurement noise on observation quality \cite{kalaitzakis2021fiducial}.
To assess the effect of range measurement noise, we analyze the ATE under different noise conditions, as depicted in Fig. \ref{simrobust}d. 
As the level of range measurement noise rises, the discrepancy between observed and actual BMAV locations widens, resulting in an increase in ATE across all methods. 
TransformLoc exhibits superior performance compared to the baselines across various noise conditions. 
By strategically planning the trajectories of AMAVs to provide observations in closer proximity to BMAVs, TransformLoc effectively mitigates the impact of noise on localization accuracy of BMAVs.
}

\begin{figure*}[t]
\setlength{\abovecaptionskip}{0.cm} 
\setlength{\belowcaptionskip}{-0.4cm} 
\setlength{\subfigcapskip}{-0.1cm}  
\centering
    \subfigure[\revise{Effectiveness of indicator}]{
        \centering
        \includegraphics[width=0.65\columnwidth]{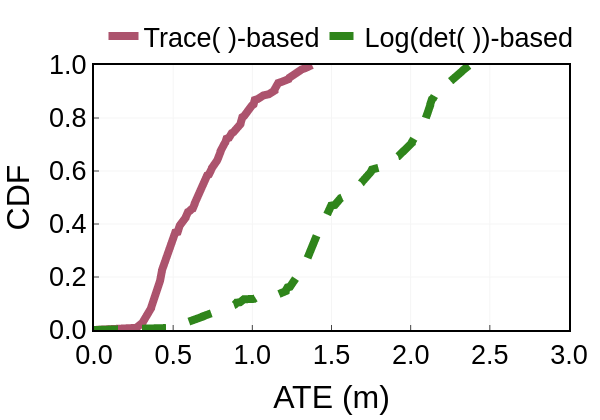}
    }
    \subfigure[\revise{Effectiveness of grouping}]{
        \centering
        \includegraphics[width=0.65\columnwidth]{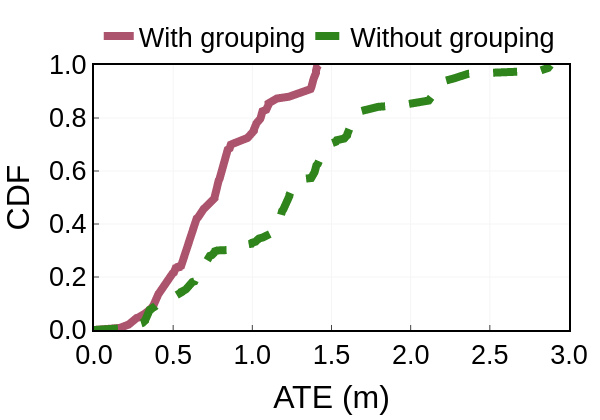}
    }
    \subfigure[\revise{Motion command interval}]{
        \centering
        \includegraphics[width=0.65\columnwidth]{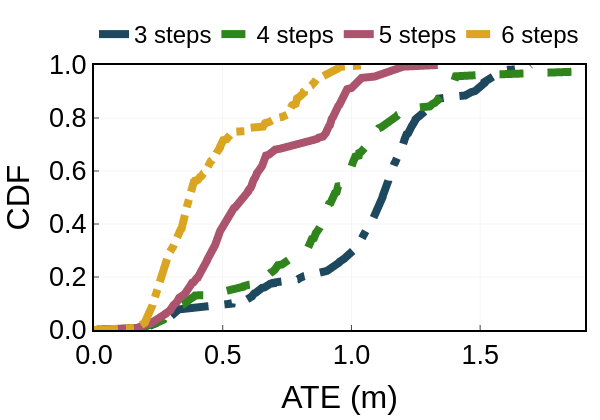}
    }

\caption{System Micro-benchmark}
\vspace{-0.4cm}
\label{Micro-benchmark}
\end{figure*}

\begin{figure*}[t]
\setlength{\abovecaptionskip}{0.cm} 
\setlength{\belowcaptionskip}{-0.4cm} 
\setlength{\subfigcapskip}{-0.1cm}  
\centering
    \subfigure[\revise{Impact of parameter}]{
        \centering
        \includegraphics[width=0.68\columnwidth]{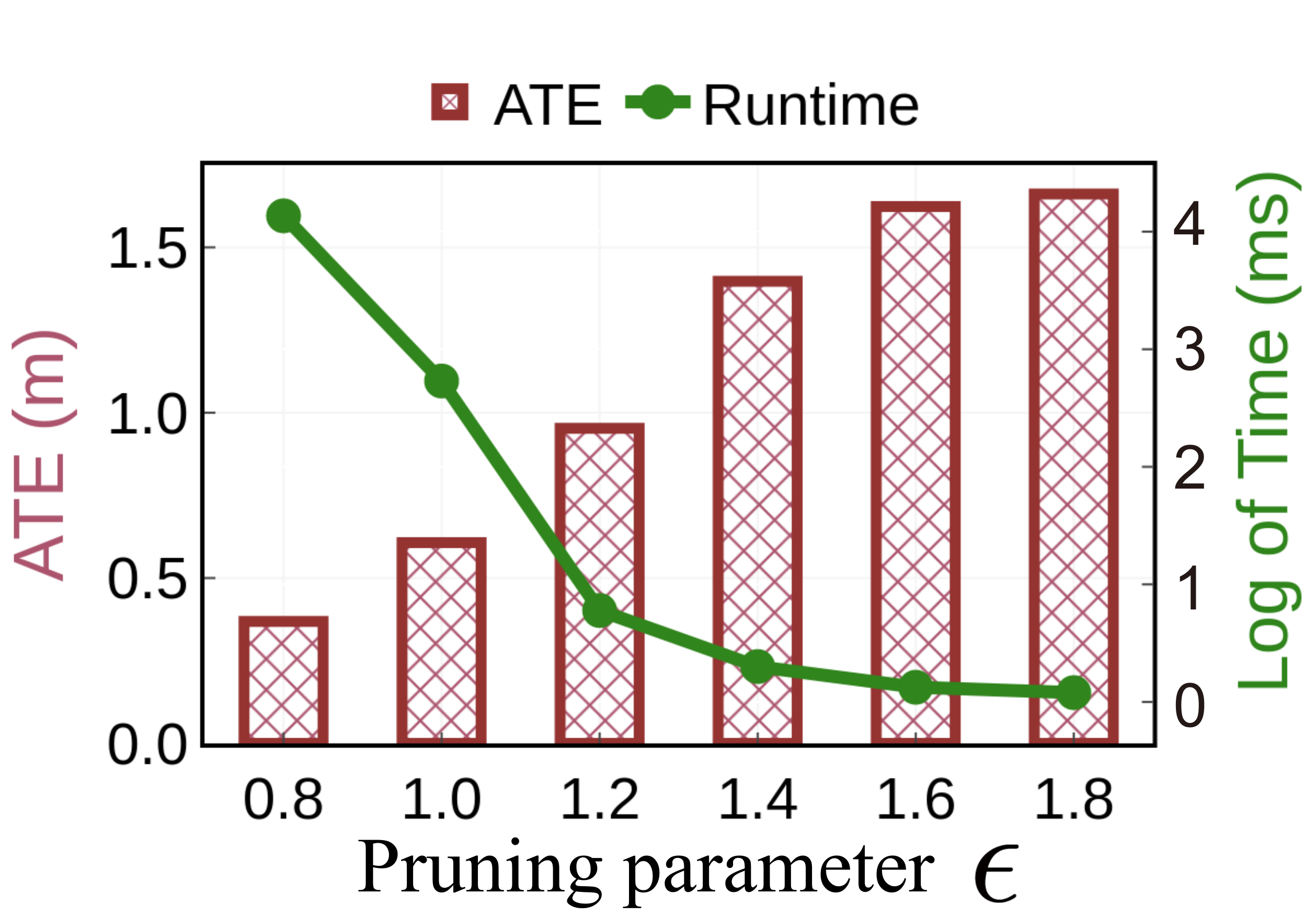}
    }
    \subfigure[Correlation of ATE and indicator]{
        \centering
            \includegraphics[width=0.60\columnwidth]{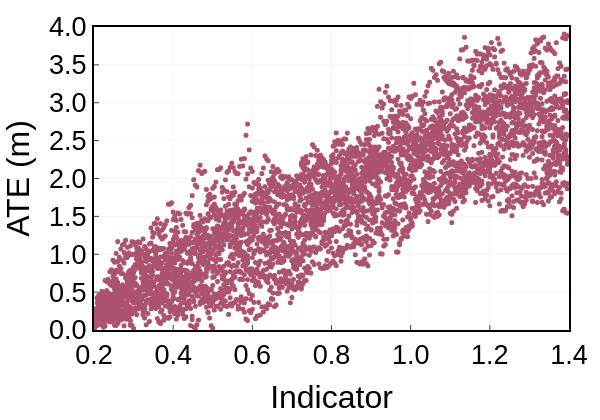}
    }
    \subfigure[\revise{Localization latency}]{
        \centering
            \includegraphics[width=0.66\columnwidth]{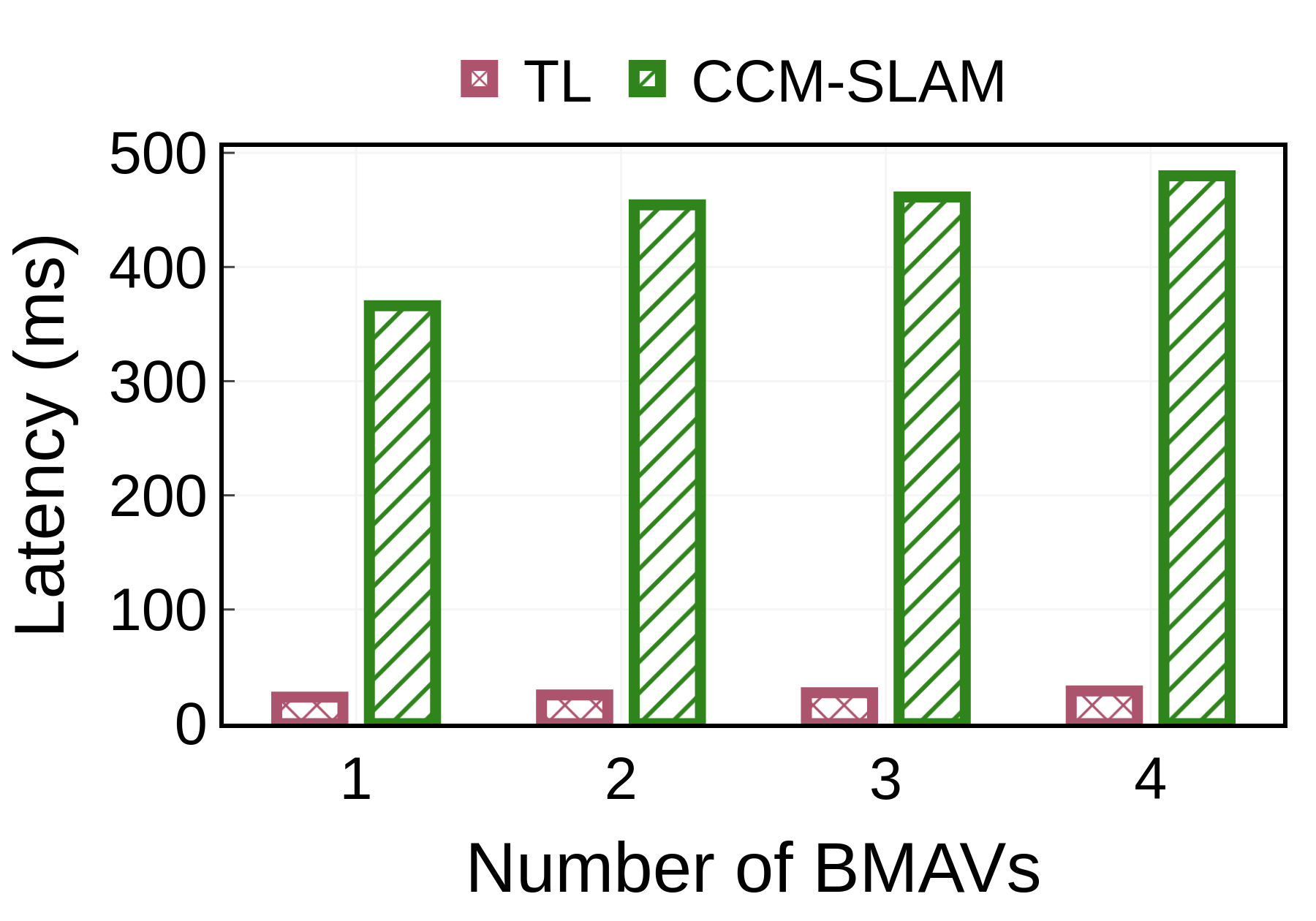}
    }
\caption{System efficiency}
\vspace{-0.4cm}
\label{efficiency}
\end{figure*}
\revise{
\subsection{System Micro-benchmark}
We conduct additional experimental analyses to scrutinize key components of TransformLoc, focusing on the individual performance enhancements contributed by each component to the overall system.

\subsubsection{Effectiveness of Indicator}
In Fig. \ref{Micro-benchmark}a, we present the CDF of the ATE using two different indicators: the trace-based and the logarithm determinant-based \cite{wang2022h}. 
The results reveal that the trace-based indicator consistently yields lower errors.
Particularly noteworthy is the sharper incline in the CDF curve based on the trace-based indicator, even for ATE values below $1.0m$, compared to the slower growth observed with the logarithm determinant-based indicator. 
This difference suggests that trace-based indicator provides a more precise estimation for BMAVs, resulting in enhanced overall performance.

\subsubsection{Effectiveness of grouping Module}
In \fig \ref{Micro-benchmark}b, we analyze the influence of the grouping Module on the overall localization performance. 
The results demonstrate that the presence of the grouping module consistently leads to lower ATE values. 
Specifically, the system incorporating the grouping module exhibits a maximum ATE lower than 1.4m, while the system without it experiences a maximum error nearly twice as large. 
These findings underscore the effectiveness of the grouping module in efficiently allocating AMAVs' sensing resources, thereby enhancing utility and ensuring a timely response for BMAVs' localization.

\subsubsection{Motion command interval of BMAVs}
In \fig \ref{Micro-benchmark}c, the CDFs of ATE are depicted for four different command intervals, demonstrating that larger intervals correspond to lower localization errors. A larger interval allows TransformLoc to anticipate the motion of BMAVs across multiple future states, providing richer information for the strategic planning of AMAVs. Consequently, this facilitates more accurate localization for BMAVs.
}

\revise{
\subsection{System Efficiency}
\subsubsection{Impact of parameter}
The AMAV search tree pruning method involves eliminating nodes in the search tree that are physically close and exhibit similar BMAV covariances. 
Evaluating $\epsilon$-algebraic redundancy constitutes the computationally intensive core operation in this pruning process. To investigate the interplay between ATE, computational time, and the pruning parameter $\epsilon$, we generate Fig. \ref{efficiency}a with varying values of $\epsilon$. 
The results illustrate that increasing $\epsilon$ leads to reduced computational time, owing to the decrease in the number of search tree nodes.
However, this reduction in time comes at the cost of increased ATE, as the policy tends towards a more greedy approach, posing challenges in achieving non-myopic objectives. 
Notably, when $\epsilon$ is set to $1.2$, the AMAV planning process requires less than $10ms$ to execute, with further reductions in time attainable by increasing $\epsilon$.

\subsubsection{Correlation of ATE and indicator}
To gauge the efficacy of the indicator, we delve deeper into the insights provided by Fig. \ref{efficiency}b. The visual representation of the indicator ($tr(\Sigma_{j,t})$) unveils a conspicuous association with the ATE of BMAVs. This elucidation underscores the indicator's capacity to faithfully mirror the degree of localization error experienced by the BMAVs, affirming its utility as a reliable metric for assessing positional accuracy.

\subsubsection{Localization latency}

To evaluate the latency of TransformLoc, we analyze the data presented in Fig. \ref{efficiency}c. The findings unveil that TransformLoc exhibits remarkable efficiency, swiftly generating observations and localizing four BMAVs within a mere $28.58ms$. This swift processing time underscores the agility of TransformLoc in handling localization tasks. In stark contrast, the comparative method, CCM-SLAM, lags significantly behind, demanding a substantially longer time of $479.93ms$ for aligning point clouds to facilitate localization. This notable discrepancy highlights the superior speed and efficiency of TransformLoc in comparison to traditional methods like CCM-SLAM.
}
\section{Related Work}\label{7}
\noindent \textbf{MAV swarm-based collaborative sensing.}
The convergence of AI advancements with the manufacturing sector has ignited a surge in research exploration within MAV swarms \cite{chen2019asc, chen2022deliversense}, unlocking augmented sensing capabilities and accelerated response rates during critical situations through their synchronized operational framework and collaborative synergy \cite{wang2021lifesaving, li2024quest, chen2024ddl}. 
These swarms are outfitted with a diverse array of sensors, spanning from cameras \cite{wang2024joint} to Lidar \cite{li2022motion}, Radar \cite{lu2020see}, IMU \cite{zhao2023smoothlander, chen2015drunkwalk}, acoustic sensors \cite{wang2022micnest}, and gas sensors \cite{shan2021ultra}, empowering them to undertake an extensive range of emergency sensing tasks across challenging environments like urban search and rescue operations \cite{rashid2020socialdrone, albanese2021sardo, xu2019ilocus}, hazardous gas detection scenarios \cite{chen2020pas, mcguire2019minimal, liu2024mobiair}, and post-disaster data transmission endeavors \cite{ren2023scheduling, kumar2019analysis, arribas2019coverage, li2021physical}. 
The data harvested by MAVs is seamlessly transmitted through communication channels \cite{xu2024emergency}, offering valuable insights into the swarm's mission objectives and the real-time status of individual MAVs \cite{alsamhi2021predictive, da2019tagging}, inclusive of their dynamic motion strategies \cite{zhou2021intelligent, 8750806, chen2024soscheduler}.
Nevertheless, the adoption of MAVs equipped with such sophisticated sensor suites entails considerable financial investment, often rendering the integration of a vast number of MAV nodes within a swarm economically impractical, thereby imposing constraints of MAV swarm applications.
\revise{
\noindent \textbf{Localization and navigation of a MAV.}
During sensing missions, the precise localization and navigation of MAVs play a pivotal role in facilitating effective collaboration within MAV swarms, especially considering the inherent limitations in computing power, communication bandwidth, and sensing capabilities of individual MAVs \cite{gowda2016tracking}. 
The adoption of SLAM techniques has become prevalent \cite{mur2015orb, davison2007monoslam, jian2024lvcp}, wherein MAVs are equipped with a variety of sensors such as RGB cameras, LiDAR, and depth cameras to gather environmental data for map construction and self-state information for localization, subsequently employing navigation algorithms to guide their movements \cite{zhou2022swarm}. 
\majorrevise{
However, these methods often demand substantial sensing and computational resources, making them applicable to AMAVs but unsuitable for BMAVs \cite{chen2017design}.
Existing methods primarily utilize and navigate AMAVs for environment mapping or exploration \cite{xu2022swarmmap, zhou2023racer}.
Specifically, these methods rely on the AMAVs' onboard sensors, such as RGB cameras, LiDAR, and depth cameras, to collect environmental data. The collected data is then analyzed to construct an environmental map or guide the AMAVs' navigation.
In contrast, our approach in TransformLoc introduces a novel concept: while AMAVs are employed to gather environmental data, their onboard sensors—equipped with advanced sensing capabilities—are also repurposed to observe and assist in localizing BMAVs, which have limited sensing and computational resources. 
Moreover, TransformLoc takes into account the localization error of BMAVs when navigating AMAVs.
}
To locate MAVs, methods based on external infrastructure are widely used. 
These approaches employ fixed infrastructure such as GPS\cite{liu2022deepgps}, microphones\cite{yimiao2023indoor, yuan2023acoustic}, Wi-Fi AP\cite{chi2024rf, zheng2019zero, guo2022wepos}, and mmWave radar\cite{iizuka2023millisign} to determine the MAV's location, offering commendable accuracy at minimal sensor expenses.
Nonetheless, these approaches often necessitate the installation of additional infrastructure and are often limited by their range, especially in obstructed environments, posing feasibility concerns in hazardous environments \cite{wang2019computer, kumar2014accurate}. 
}

\revise{
\noindent \textbf{Localization and navigation of a MAV swarm.}
Collaborative SLAM methodologies, explored in prior research endeavors, face significant challenges due to the prohibitive costs of the required sensors, thereby constraining the size of the swarm \cite{jian2023path}. 
For instance, \cite{xu2022swarmmap} introduced SwarmMap, a framework designed for multi-robot systems to collaboratively build and share maps in real-time, demonstrating the potential of distributed SLAM in improving efficiency and accuracy in navigation tasks. 
Similarly, \cite{xu2020edge} presented Edge-SLAM, which leverages edge computing resources to offload intensive SLAM computations from the MAVs, enhancing their operational efficiency and battery life.
Despite these advancements, the high cost and resource demands of the sensors required for these systems remain a significant barrier, as highlighted by \cite{schmuck2019ccm}, who noted that the cost of sensors such as LiDAR and high-resolution cameras can limit the deployment of large-scale swarms. 
This financial constraint not only restricts the number of MAVs in a swarm but also impacts the overall feasibility and scalability of collaborative SLAM systems in real-world applications. 
Consequently, there is a pressing need for more cost-effective solutions that can facilitate large-scale swarm deployments without compromising on performance.
}

\majorrevise{
In response, we propose TransformLoc, a collaborative and adaptable localization framework tailored for heterogeneous MAV swarms. 
Within TransformLoc, a select number of AMAVs serve as mobile localization infrastructure for a multitude of BMAVs, thereby enabling precise localization at a reduced overall cost. 
TransformLoc builds upon existing localization techniques by enabling accurate AMAV localization and leveraging external observations from AMAVs to assist BMAV localization.
Therefore, TransformLoc's adaptive grouping and scheduling strategy is fundamentally grounded in a broader range of existing localization techniques.
Furthermore, any localization technique designed for individual MAVs or homogeneous MAV swarms can be integrated into TransformLoc to enhance the accuracy of AMAV or BMAV localization, thereby improving the overall localization performance of the heterogeneous swarm.
In terms of localization methods for heterogeneous UAV swarms, approaches like H-DrunkWalk \cite{chen2020h} establish a temporary infrastructure by designating a few landed MAVs as radio beacons. These beacons emit radio signatures that allow the algorithm to detect trajectory intersections of mobile MAVs. By leveraging these intersections, the algorithm integrates noisy dead-reckoning measurements from multiple MAVs to enhance their location accuracy. However, this method relies heavily on the self-established infrastructure, which reduces the operational efficiency of the swarm. TransformLoc repurposes a select number of AMAVs as mobile localization infrastructures, redistributing their sensing and computational resources to improve the location estimation of a larger fleet of BMAVs.
This innovative solution not only enhances the scalability and affordability of MAV swarm applications but also broadens their applicability across diverse operational scenarios.
}

\section{Discussion} \label{8}

In this part, we explore several influential factors affecting TransformLoc's operation.\\
$(i)$ \textit{Communication Overload.}
BMAVs regularly transmit location estimations (mean and covariance) to the AMAVs. With twenty BMAVs, each AMAV processes and transmits less than 10KB of data every $\delta$ seconds. 
Similarly, during grouping, an AMAV gathers location information from other AMAVs, resulting in a total data amount of less than 10KB with twenty AMAVs. Thus, the communication load remains within acceptable limits.\\
$(ii)$ \textit{Localization accuracy of AMAV.}
When the localization error of an AMAV is less than 10$cm$, the system significantly enhances the localization accuracy of BMAVs. 
Achieving this level of accuracy is feasible with an AMAV equipped with advanced sensors like depth cameras. \\
\majorrevise{
$(iii)$ \textit{The quantity of AMAV and BMAV.}
In the best-case scenario, the number of BMAVs equals the number of AMAVs ($N=M$). 
Under this condition, each AMAV can consistently observe a single BMAV, transforming its sensing and computational resources to it, thereby maintaining the swarm's localization error at a relatively low level. 
As the ratio of BMAVs to AMAVs ($N/M$) increases, the spatial dispersion of BMAVs grows, leading to some BMAVs always falling outside the AMAVs’sensing range, which can degrade localization accuracy. Speed limitations exacerbate this issue: AMAVs may struggle to track dispersed BMAVs, while AMAVs waste energy maneuvering without effectively reducing localization errors. Additionally, the limited FOV of AMAVs restricts their ability to observe multiple BMAVs simultaneously, particularly in dense or scattered configurations, resulting in unmonitored regions and increased localization error. These factors collectively undermine the swarm's accuracy and stability. Our experiments in an $8m$ $\times$ $8m$ environment show that when the number of AMAVs is 5, the localization error increases significantly if the number of BMAVs exceeds 25, with a notable rise in collision risk. Thus, the worst-case scenario occurs when the number of BMAVs is at least five times that of AMAVs ($N \geq 5M$), where both error and collision risk escalate.\\
$(iv)$ \textit{Scalability of TransformLoc.}
When further enhancing the scalability of TransformLoc to larger swarms or more complex environments \cite{singhal2017resource}, several factors need to be considered: 
(1) Resource Allocation Efficiency. The adaptive grouping-scheduling strategy has demonstrated effectiveness in allocating sensing and computing resources of AMAVs within moderate-sized swarms. For larger swarms, techniques such as hierarchical grouping or distributed scheduling could enhance scalability. Hierarchical grouping organizes MAVs into nested subgroups, allowing localized decision-making and reducing the burden on central coordination \cite{arafat2019localization}. Distributed scheduling, on the other hand, leverages decentralized algorithms to enable individual MAVs or swarms to independently manage their tasks, thereby enhancing responsiveness \cite{zhou2023racer}. 
Additionally, learning-based grouping-scheduling~\cite{ wang2023gcrl}, which leverages attention mechanisms to model the dynamic relationships among MAVs or swarms, offers another promising approach for adaptive grouping and allocation efficiency enhancing.
(2) Computational Overhead. Increasing the number of MAVs introduces higher computational demands for joint location estimation and scheduling. To address these challenges, strategies such as pruning in the search tree and parallel processing can be highly effective. Pruning in the search tree reduces computational overhead by eliminating suboptimal or redundant branches early, thereby focusing resources on promising solutions \cite{atanasov2014information}. Parallel processing, meanwhile, leverages the computational power of multi-core processors or distributed systems to execute location estimation and scheduling tasks simultaneously, dramatically improving efficiency. Additionally, hybrid approaches that integrate heuristic optimization algorithms or learning-based methods, such as reinforcement learning \cite{jeong2019learning}, can further enhance scalability and adaptability, especially in dynamic or resource-constrained environments.
(3) Environmental Complexity. In more dynamic or complex environments, additional constraints such as obstacle avoidance or varying communication reliability must be addressed. The modular design of TransformLoc allows for the integration of corresponding solutions into the framework to address these constraints. For example, advanced obstacle avoidance algorithms for both AMAVs and BMAVs \cite{zhou2022swarm, duisterhof2021sniffy} can be embedded into the framework, leveraging real-time sensor data and predictive modeling to enable safe navigation in cluttered or rapidly evolving environments. 
To enhance scalability, adaptive and efficient communication media are key. Wi-Fi mesh networks expand coverage dynamically \cite{ashraf2014capacity}, LoRa supports large-scale deployments with low-power, long-range communication \cite{soorki2024catch}, and 5G leverages massive device connectivity and network slicing for scalability enhancement \cite{bertizzolo2021streaming}. Combining Wi-Fi for bandwidth with LoRa for coverage could ensure scalable performance in complex environments. Furthermore, implementing more stable and adaptive communication methods, such as multi-hop relay protocols, can enhance connectivity and reduce latency in scenarios where communication reliability fluctuates \cite{singhal2017resource}.
\\
$(v)$ \textit{Time complexity.}
TransformLoc consists of two main components: Error-Aware Joint Location Estimation and Similarity-Instructed Adaptive Grouping-Scheduling.
(1) Error-Aware Joint Location Estimation. 
In the Prediction Step, the algorithm performs two operations: state propagation through the motion model and covariance propagation, with respective time complexities $O(n^2)$ and $O(n^3)$, where $n$ represents the state dimension.
These complexities arise due to matrix multiplication and inversion. For $N$ BMAVs, the total time complexity of this step is $O(Nn^3)$. 
The BMAV state includes position information, denoted as $y_{i, t}$.
In the Correction Step, the operations include checking if a BMAV is within an AMAV's FoV, observation model updates, and Kalman corrections. For $N$ BMAVs and $M$ AMAVs, these operations have respective time complexities of $O(NM)$, $O(n^2)$, and $O(n^3)$. 
Thus, the time complexity of this step is $O(NMn^3)$.
(2) Similarity-Instructed Adaptive Grouping-Scheduling.
In the Graph-Based Adaptive Grouping, the operations include computing Voronoi diagrams for AMAV grouping and assigning BMAVs to AMAV regions. For $N$ BMAVs and $M$ AMAVs, these operations have time complexities of $O(MlogM)$ and $O(N)$, respectively, resulting in a total time complexity of  $O(MlogM + N)$ for this step.
In the Search Tree-Based Non-Myopic Scheduling, the operations involve constructing the search tree and pruning nodes. For each AMAV with $k$ motion options, a lookahead horizon of $\delta$, and $p$ pruned nodes, these operations have time complexities of $O(k^\delta Nn^2)$ and $O(pNn^2)$, respectively. The overall time complexity of this step is $O(Mk^\delta Nn^2)$.
The total time complexity of the algorithm is $O(NMn^3 + Mk^\delta Nn^2 + MlogM + N)$. From this, we derive several key insights: 1. High-Dimensional State ($n$): The $n^3$ cost, arising from Kalman updates, dominates in this case. 2. Large Control Space or Horizon ($k, \delta$): The  $k, \delta$ factor becomes dominant in the scheduling step, especially with many AMAVs ($M$). Tree pruning is essential for ensuring real-time feasibility. 3. Number of Agents ($N, M$): The  factor $NM$ significantly affects the correction step and AMAV-BMAV pairing. Larger swarms increase computational demands depending on the component.
}

\section{Future work} \label{9}
In this part, we discuss potential future research directions:\\
$(i)$ \textit{Enhancing system robustness to AMAV localization errors.} Currently, our system uses tracking cameras for AMAV localization, which provide high precision. However, in real-world environments, AMAVs may encounter increased localization errors. Improving TransformLoc's tolerance to these errors and the precision of observations generated by AMAVs for BMAVs is a key area for future work, and utilizing the spatio-temporal relationship between MAVs is a feasible solution \cite{wang2023califormer}.\\
$(ii)$ \textit{Scaling up the swarm.} In our current tests, TransformLoc supports 7 AMAVs and 28 BMAVs. To extend the system's operational range, we need to accommodate more MAVs. Expanding the number of supported AMAVs and BMAVs is one of our next research directions.\\
$(iii)$ \textit{Faster resource allocation for AMAVs.} Presently, we use a tree-based scheduling approach for non-myopic resource allocation among AMAVs. As the search depth increases, this may introduce greater delays. Reducing the resource allocation delay for AMAVs is another area for research.\\
\majorrevise{
$(iv)$ \textit{A more comprehensive observation noise model.} 
In our observation model, the assumption of a zero-mean error was made to emphasize the system's core functionalities. However, real-world environmental factors, such as varying illumination, sensor noise, and occlusions, can introduce biases or result in non-zero mean errors in practical scenarios. To address these influences, a more comprehensive observation noise model could be developed by incorporating additional parameters or empirical data. \\
$(v)$ \textit{Varying altitudes and diverse environmental conditions.} 
Our framework , TransformLoc, is able to run when AMAVs and BMAVs have different altitudes, and has the ability to run in diverse operational conditions.
The TransformLoc framework assumes constant altitude and equal operational conditions to focus on the core contributions of TransformLoc. 
These assumptions allow us to systematically evaluate the effectiveness of our approach under controlled conditions. 
However, real-world scenarios may feature varying altitudes and diverse environmental conditions for both AMAVs and BMAVs.
To address this, TransformLoc's modular design provides the flexibility to incorporate additional constraints or adaptations. 
For example: 1. Varying Altitudes: The framework can be extended to include a 3D model of MAV dynamics and resource allocation strategies to account for vertical motion and observation overlaps. 2. Diverse environmental conditions: TransformLoc can integrate contextual factors such as weather, terrain, or MAV-specific capabilities to dynamically adjust its estimation and scheduling strategies. 
}

\section{Conclusion} \label{10}
In this work, we introduce TransformLoc, a novel framework that revolutionizes the role of AMAVs by transforming them into mobile localization infrastructures. TransformLoc is distinguished by two key innovations: Firstly, we introduce an error-aware joint location estimation model. This model seamlessly integrates the imprecise estimates from BMAVs with observations from AMAVs, facilitating joint location estimation of BMAVs. To achieve this, we employ an uncertainty-aided inference method, ensuring robustness and accuracy in localization. Secondly, we introduce a similarity-instructed adaptive grouping-scheduling strategy. This innovative approach dynamically allocates AMAVs' sensing resources through a combination of a graph-based adaptive grouping method and a search tree-based non-myopic scheduling method. Additionally, we propose a similarity-instructed search tree pruning method to further improve computational efficiency. Experimental evaluations demonstrate that TransformLoc outperforms existing solutions, showcasing its efficacy and superiority.

\ifCLASSOPTIONcompsoc
  \section*{Acknowledgments}
\else
  \section*{Acknowledgment}
\fi

This paper was supported by the National Key R\&D program of China No. 2022YFC3300703, the Natural Science Foundation of China under Grant No. 62371269. Guangdong Innovative and Entrepreneurial Research Team Program No. 2021ZT09L197, and Tsinghua Shenzhen International Graduate School Cross-disciplinary Research and Innovation Fund Research Plan No. JC20220011.


\bibliographystyle{IEEEtran}
\bibliography{infocom}

\vspace{-2cm}
\begin{IEEEbiography}
[{\includegraphics[width=1.2in,height=1.2in,clip,keepaspectratio]{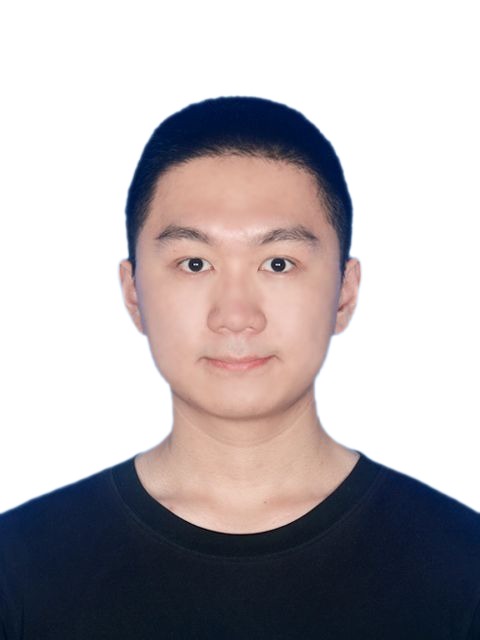}}]{Haoyang Wang} received the B.E. degree from the School of Computer Science and Engineering, Central South University, China, in 2022. He is currently pursuing the Ph.D. degree at the Tsinghua Shenzhen International Graduate School, Tsinghua University, China. His research interests include mobile computing, AIoT and distributed \& embedded AI.
\end{IEEEbiography}
\vspace{-1.5cm}
\begin{IEEEbiography}
[{\includegraphics[width=1.1in,height=1.1in,clip,keepaspectratio]{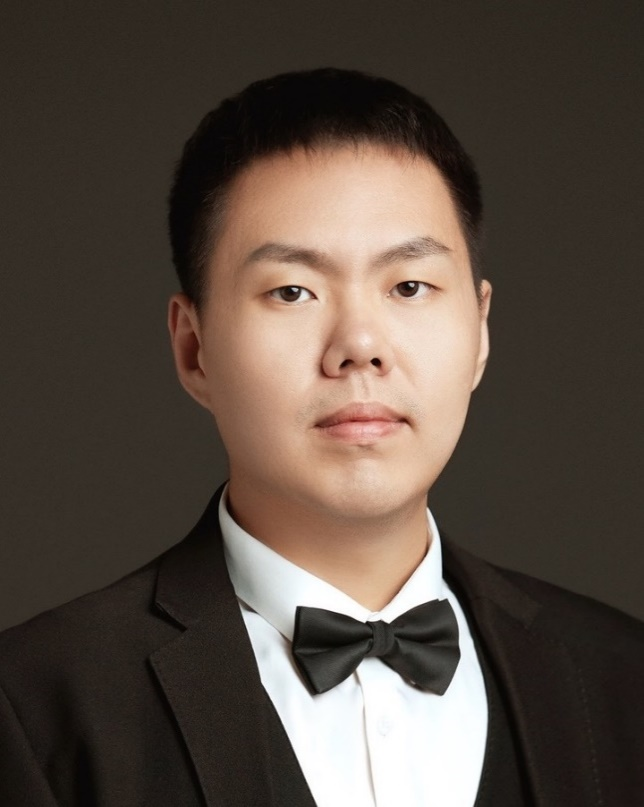}}]{Jingao Xu} is now a postdoctoral researcher in School of Software, Tsinghua University. He received both his Ph.D. and B.E. degrees from Tsinghua University in 2022 and 2017, respectively. His research interests include Internet of Things, mobile computing, and edge computing.
\end{IEEEbiography}
\vspace{-1.5cm}
\begin{IEEEbiography}
[{\includegraphics[width=1.3in,height=1.25in,clip,keepaspectratio]{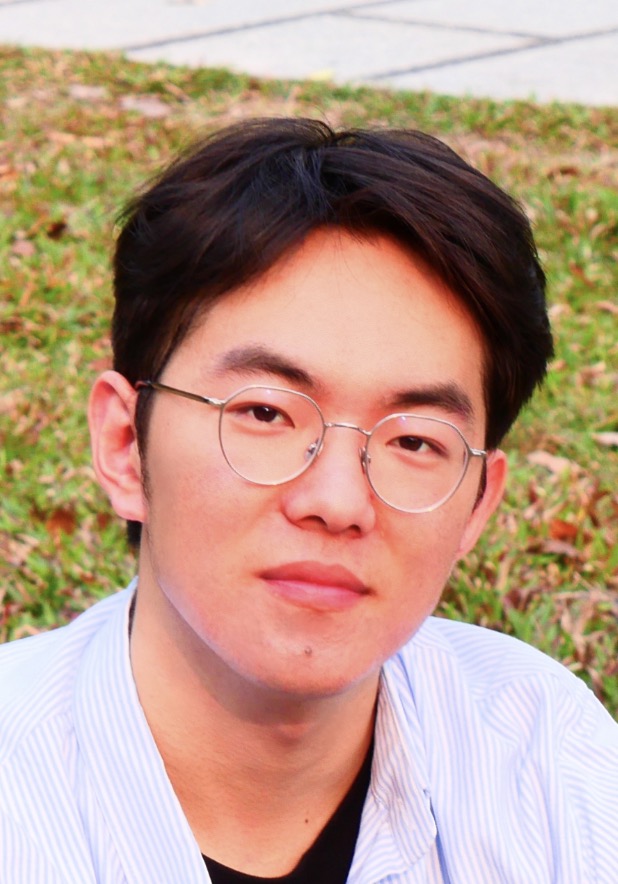}}]{Chenyu Zhao} received the B.E. degree of Electronic Information Science and Technology from Northwest University in China, and the B.S. degree of Electronic System Engineering from University of Essex in UK, in 2022. He is currently pursuing the Master degree at the Tsinghua Shenzhen International Graduate School, Tsinghua University, China. His research interests include Mobile Computing, Cyber-Physical Systems, Robotics, and Quadrotors.
\end{IEEEbiography}
\vspace{-1.5cm}
\begin{IEEEbiography}
[{\includegraphics[width=1.3in,height=1.25in,clip,keepaspectratio]{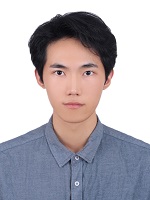}}]{Yuhan Cheng} received his B.E. degree from the School of Intelligent Systems Engineering at Sun Yat-Sen University, Guangzhou, China, in 2022. He is currently pursuing an M.S. degree at Shenzhen International Graduate School, Tsinghua University, Shenzhen, China. His research interests encompass cyber-physical systems and collaborative multi-agent robotic motion scheduling.
\end{IEEEbiography}
\vspace{-1.5cm}
\begin{IEEEbiography}
[{\includegraphics[width=1.3in,height=1.25in,clip,keepaspectratio]{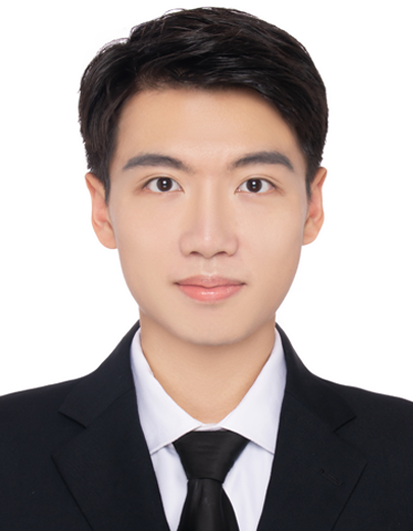}}]{Xuecheng Chen} received the B.E. degree from the Department of Intelligence Engineering, Sun Yat-sen University, Guangzhou, China, in 2021. He is currently pursuing the Ph.D. degree at the Department of Tsinghua-Berkeley Shenzhen Institute (TBSI), Tsinghua University, Shenzhen, China. His current research interests include mobile computing, multi-robot systems, and cyber-physical systems.
\end{IEEEbiography}
\vspace{-1cm}
\begin{IEEEbiography}
[{\includegraphics[width=1.3in,height=1.25in,clip,keepaspectratio]{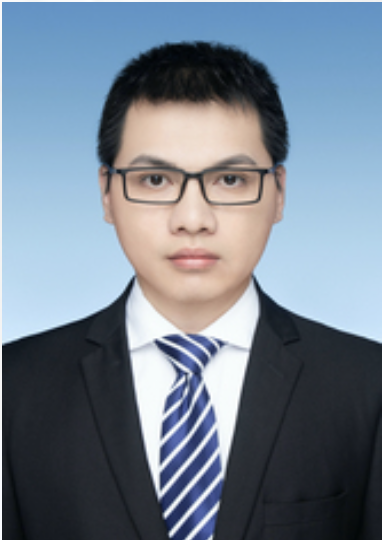}}]{Chaopeng Hong} received his B.E. and Ph.D. degrees from Tsinghua University in 2012 and 2017, respectively. He is currently an Assistant Professor at Tsinghua Shenzhen International Graduate School, Tsinghua University, Shenzhen, China. His research interests lie in the coupled human-environment systems and sustainable systems analysis, including especially climate change, air quality, agriculture, energy and water systems
\end{IEEEbiography}

\begin{IEEEbiography}
[{\includegraphics[width=1.3in,height=1.25in,clip,keepaspectratio]{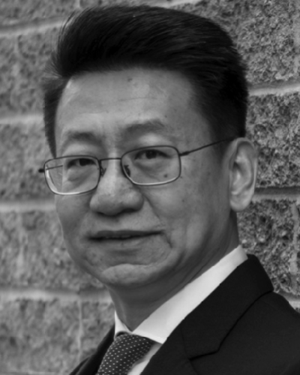}}] {Xiao-Ping Zhang} (Fellow, IEEE) received the B.S. and Ph.D. degrees in electronic engineering from Tsinghua University, Beijing, China, in 1992 and 1996, respectively. 
He is the Chair Professor at Tsinghua-Berkeley Shenzhen Institute (TBSI), Tsinghua University, Shenzhen, China. Since Fall 2000, he has been with the Department of Electrical, Computer and Biomedical Engineering, Ryerson University, Toronto, ON, Canada, where he is currently a Professor and the Director of the Communication and Signal Processing Applications Laboratory. In 2015 and 2017, he was a Visiting Scientist with the Research Laboratory of Electronics, Massachusetts Institute of Technology, Cambridge, MA, USA. His research interests include sensor networks and the Internet of Things (IoT), machine learning, statistical signal processing, image and multimedia content analysis, and applications in big data, finance, and marketing. Dr. Zhang is a fellow of the Canadian Academy of Engineering and the Engineering Institute of Canada. He was a recipient of the 2020 Sarwan Sahota Ryerson Distinguished Scholar Award and the Ryerson University Highest Honor for scholarly, research, and creative achievements. He was selected as the IEEE Distinguished Lecturer by the IEEE Signal Processing Society from 2020 to 2021 and the IEEE Circuits and Systems Society from 2021 to 2022.
\end{IEEEbiography}
\vspace{-0.5cm}
\begin{IEEEbiography}
[{\includegraphics[width=1.3in,height=1.25in,clip,keepaspectratio]{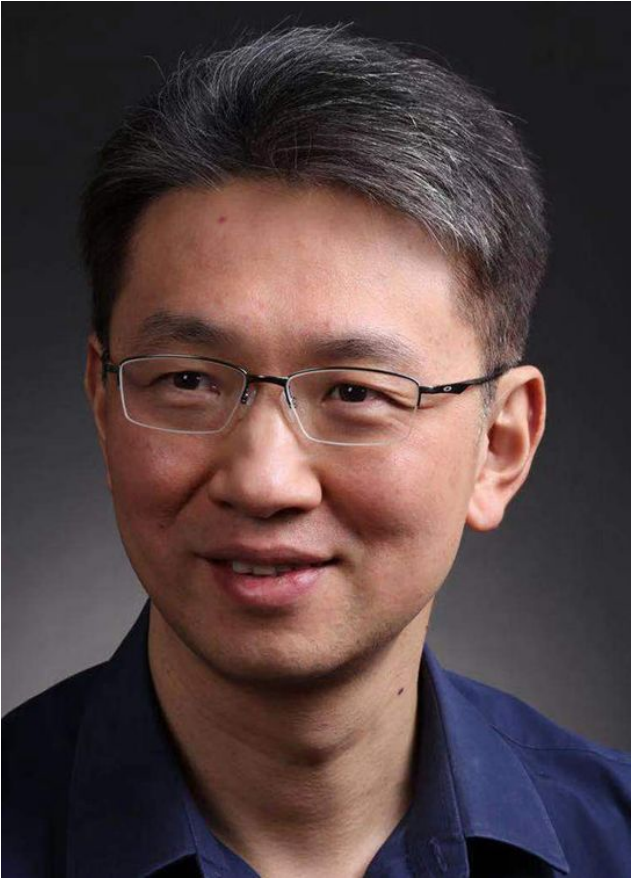}}]{Yunhao Liu} (Fellow, IEEE) received his B.E. degree from the Department of Automation, Tsinghua University, Beijing, in 1995. He received his M.A. degree from Beijing Foreign Studies University, Beijing, in 1997. He received his M.S. and Ph.D. degrees in computer science and engineering from Michigan State University, East Lansing, in 2003 and 2004, respectively. He is a professor in the Department of Automation and the dean of the Global Innovation Exchange, Tsinghua University, Beijing. He is a fellow of CCF, ACM and IEEE. His research interests include Internet of Things, wireless sensor networks, indoor localization, the Industrial Internet, and cloud computing.
\end{IEEEbiography}
\vspace{-0.5cm}
\begin{IEEEbiography}
[{\includegraphics[width=1.3in,height=1.25in,clip,keepaspectratio]{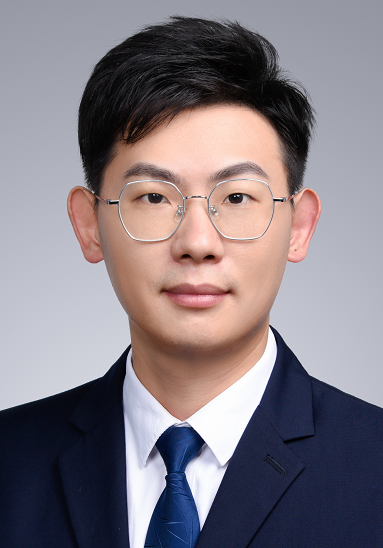}}]{Xinlei Chen} (Member, IEEE) received the B.E. and M.S. degrees in electronic engineering from Tsinghua University, China, in 2009 and 2012, respectively, and the PhD degrees in electrical engineering from Carnegie Mellon University, Pittsburgh, Pennsylvania, in 2018. 

He was a postdoctoral research associate in Electrical Engineering Department, Carnegie Mellon University, Pittsburgh, Pennsylvania. He is currently an Assistant Professor with Shenzhen International Graduate School, Tsinghua University, Shenzhen, Guangdong, China. He is also with Pengcheng Lab, Shenzhen, Guangdong, China and RISC-V International Open Source Laboratory, Shenzhen, Guangdong, China. His research interests include AIoT, artificial intelligence, pervasive computing, cyber physical system, robotics, urban sensing, brain computer interface and human computer interface.
\end{IEEEbiography}

\end{document}